\documentclass[12pt]{iopart}

\pdfoutput=1

\usepackage{color}
\usepackage{cite}
\usepackage{url,psfrag,graphicx}
\usepackage{dcolumn}
\usepackage{bm}
\usepackage{hyperref}
\usepackage{float,epsfig,color}
\usepackage{times}
\usepackage{array}
\usepackage{braket}
\usepackage{diagbox}
\usepackage{lscape}
\usepackage{multirow}
\usepackage{stackrel}
\usepackage{verbatim}
\usepackage{xcolor}
\usepackage{soul}
\usepackage{hyperref}
\usepackage{iopams}
\usepackage{setstack}

\newcommand{\eqref}[1]{(\ref{#1})}

\DeclareMathAlphabet      {\mathitbf}{OML}{cmm}{b}{it}

\definecolor{ForestGreen}{RGB}{34,139,34}

\begin{document}

\title[Discrete Lattice Models for Interface Growth on a Complete Graph]{Discrete Lattice Models for Interface Growth on a Complete Graph}

\author{J. M. Marcos$^{1,2}$, J. J. Mel\'endez$^{1,2}$, R. Cuerno$^3$, J. J. Ruiz-Lorenzo$^{1,2}$}

\address{$^1$ Departamento de  F\'{\i}sica, Universidad de Extremadura, 06006 Badajoz, Spain}
\address{$^2$ Instituto de Computaci\'on Cient\'{\i}fica Avanzada de Extremadura (ICCAEx), Universidad de Extremadura, 06006 Badajoz, Spain}
\address{$^3$ Universidad Carlos III de Madrid, Departamento de Matemáticas and Grupo Interdisciplinar de Sistemas Complejos (GISC), Avenida de la Universidad 30, 28911 Leganés, Spain}

\date{\today}

\begin{abstract}

We investigate the behavior of discrete interface growth models belonging to the Edwards--Wilkinson (EW) and Kardar--Parisi--Zhang (KPZ) universality classes, when defined on a complete graph, a topology commonly used to probe the infinite-dimensional limit of statistical mechanical systems. Our aim is to assess to what extent discrete lattice models reproduce the behavior of their corresponding continuum equations in this highly connected setting. After assessing the trivial behavior shown by some well known cases (like random deposition with surface relaxation or the etching model), we focus on two paradigmatic models associated with the KPZ universality class, the Restricted Solid-on-Solid (RSOS) and Ballistic Deposition (BD) models, and assess non-trivial behavior through several observables including the roughness, height fluctuations, power spectra, and two-time autocorrelation functions. 
Still, despite similarities with continuum equations, important differences arise in the fluctuations and long-time dynamics. In both discrete models the distribution of rescaled height fluctuations display a pronounced left tail, indicating the presence of lagging nodes. While the RSOS model exhibits a saturation roughness that decreases with system size, similarly to the EW and KPZ equations, at intermediate times the BD model exhibits an ultrafast growth regime that is followed at long times by a time-independent roughness that increases with system size, both in contrast with the behavior of the KPZ equation on a complete graph.


\end{abstract}

\section{Introduction}\label{sec:intro}

The Kardar--Parisi--Zhang (KPZ) equation~\cite{Kardar1986} constitutes a central theoretical model for capturing the universal, scale-invariant behavior emerging in a wide variety of non-equilibrium growth processes~\cite{Barabasi1995,Krug1997,Takeuchi2018}. It describes the time evolution of a fluctuating interface and can be written as
\begin{equation}
    \frac{\partial h}{\partial t}(\boldsymbol{x},t)
    = \nu \nabla^2 h(\boldsymbol{x},t)
    + \frac{\lambda}{2} \left[ \nabla h(\boldsymbol{x},t) \right]^2
    + \eta(\boldsymbol{x},t)\, .
    \label{eq:KPZ_orig}
\end{equation}
Here, $h(\boldsymbol{x},t)$ represents the height of the interface above a reference substrate at spatial position $\boldsymbol{x} \in \mathbb{R}^d$ and time $t$. The diffusive term proportional to $\nu$ encodes relaxation effects driven by surface tension, whereas the nonlinear term with coupling constant $\lambda$ reflects lateral growth along the local normal direction of the interface. The noise term $\eta(\boldsymbol{x},t)$ accounts for stochastic fluctuations and is taken to be Gaussian white noise with vanishing mean, $\langle \eta(\boldsymbol{x},t) \rangle = 0$, and variance $2D$, such that
\begin{equation}
\label{eq:ruido}
\langle \eta(\boldsymbol{x},t)\,\eta(\boldsymbol{x}',t') \rangle
= 2D\,\delta^d(\boldsymbol{x} - \boldsymbol{x}')\,\delta(t - t')\,,
\end{equation}
where $\delta^d(\boldsymbol{u})$ denotes the $d$-dimensional Dirac delta.

Although the KPZ equation was initially formulated to model the growth of kinetically rough interfaces, its scope has broadened substantially over the past decades. It is now widely acknowledged as an effective description across diverse domains of nonequilibrium physics, including active matter systems~\cite{Caballero2020} and quantum many-body dynamics~\cite{Sieberer2025}. Moreover, this stochastic evolution equation is intimately related to other continuum field theories associated with different universality classes. In particular, in the absence of the nonlinear term, $\lambda = 0$, the KPZ equation reduces to the Edwards--Wilkinson (EW) equation~\cite{Edwards1982,Barabasi1995}, which can be interpreted as the Gaussian approximation of the stochastic time-dependent Ginzburg--Landau equation at criticality~\cite{Kardar2007}.

The scale-invariant behavior of these stochastic growth equations is conveniently summarized by the Family--Vicsek (FV) dynamic scaling hypothesis for the interface width (or roughness) $w(L,t)$~\cite{Barabasi1995,Krug1997}, see Eq.~\eqref{eq:width},
\begin{equation}
	\label{eq:w}
	w(L,t)=t^{\beta} f\!\left(t/L^z\right),
\end{equation}
where $f(\cdot)$ is a universal scaling function. This form implies an early-time power-law growth of the roughness, $w \sim t^{\beta}$ for $t \ll L^z$, followed by saturation at long times, $w_{\mathrm{sat}} \sim L^{\alpha}$ for $t \gg L^z$. The exponent $\alpha$ characterizes the amplitude of height fluctuations at large length scales, $\beta$ controls the temporal growth of the interface width, and $z$ is the dynamic exponent governing the growth of the lateral correlation length, $\xi(t) \sim t^{1/z}$. Within the FV framework, these exponents are related by the scaling identity $\alpha = \beta z$~\cite{Barabasi1995,Krug1997}.

For the EW equation, the scaling exponents can be determined exactly in any spatial dimension $d$, yielding $\alpha=(2-d)/2$, $\beta=(2-d)/4$, and $z=2$~\cite{Barabasi1995,Kardar2007}. The FV scaling holds for $d < 2$, with logarithmic corrections at the marginal dimension $d=2$, while for $d>2$ the interface becomes asymptotically flat, identifying $d^{\rm EW}_u=2$ as the upper critical dimension of the EW universality class.

For the KPZ equation, exact values of the critical exponents are known only in one space dimension, where $\alpha=1/2$, $\beta=1/3$, and $z=3/2$~\cite{Barabasi1995,Krug1997,Takeuchi2018}. In general, the KPZ fixed point is expected to satisfy the Galilean invariance scaling relation $\alpha+z=2$ for all $d\leq d^{\rm KPZ}_u$, implying that only one exponent is independent in this regime. For dimensions $d>1$, however, no exact analytical expressions for the critical exponents are available. Their determination relies on numerical integration of the KPZ equation or on simulations of discrete growth models within the same universality class; see, for instance, Refs.~\cite{Barabasi1995,Alves2014,Takeuchi2018,Oliveira2022,HalpinHealy2025} and other therein. Despite this substantial body of work, the existence and the precise value of the upper critical dimension $d^{\rm KPZ}_u$ for the KPZ universality class remain unresolved. Analytical approaches lead to disparate conclusions: some studies predict $d^{\rm KPZ}_u\le4$~\cite{Halpin-Healy1990,Bouchaud1993,Doherty1994,Colaiori2001}, others propose $d^{\rm KPZ}_u>4$~\cite{Kloss2012,Kloss2014,Kloss2014-2,Canet2025}, and yet other analyses argue for the absence of a finite upper critical dimension so that $d^{\rm KPZ}_u=\infty$~\cite{Castellano1998,Castellano1998-2}.

Alternatively to the continuum equations, there exists a wide variety of discrete lattice growth models whose large-scale, long-time behavior falls into the KPZ and EW universality classes \cite{Barabasi1995,Krug1997}. These models, defined by local stochastic rules, frequently reproduce the same scaling exponents and fluctuation statistics as their corresponding field theories \cite{Takeuchi2018}, despite having very different microscopic dynamics. This highlights that, far from equilibrium, universality is also largely governed, not by microscopic details, but by symmetries, conservation laws, and by the presence or absence of nonlinearities and noise correlations \cite{Tauber2014}, making discrete models powerful tools for both numerical and theoretical exploration of nonequilibrium surface growth phenomena.

Prior to the present study, it was established in Ref.~\cite{Marcos2026} that, when the KPZ equation is formulated on a complete graph with $N$ vertices—a substrate commonly employed to explore the infinite-dimensional limit of many statistical-mechanical systems—the KPZ nonlinearity becomes irrelevant as $N\to\infty$, giving rise to effective EW-like dynamics and asymptotically flat interfaces with vanishing roughness in the thermodynamic limit. In the same work, several theoretical predictions were also derived for relevant observables of the EW equation, providing a detailed characterization of the system behavior on a fully connected graph.

Complementary investigations have explored these continuum equations~\cite{Marcos2025}, as well as related discrete growth models~\cite{Saberi_2013,Oliveira2021}, on Cayley trees with the same goal of understanding high-dimensional or mean-field-like behavior. However, the results obtained on such substrates are strongly influenced by boundary effects. In particular, the lack of translational invariance of Cayley trees leads to interfaces that are non-flat on average~\cite{Oliveira2021}. This issue is absent in complete graphs, where all nodes are equivalent and the substrate is homogeneous by construction, making them a natural framework for investigating the infinite-dimensional limit. In this context, complete graphs were already identified in Ref.~\cite{Oliveira2021} as prototypical substrates, and standard growth models such as the Random Deposition with Surface Relaxation (RDSR) and Restricted Solid-on-Solid (RSOS) models have been shown to exhibit perfect layer-by-layer growth when defined on them.


Motivated by these results, in this paper we investigate discrete growth models in the EW and KPZ universality classes defined on a complete graph. Our aim is to determine how the dynamical behavior of representative (paradigmatic) lattice models in these classes is modified when spatial structure is replaced by all-to-all connectivity. In particular, we perform extensive numerical simulations of the RSOS and the Ballistic Deposition (BD) models on a complete graph and analyze the resulting scaling properties. Whether the phenomenology exhibited by discrete models on non-regular networks faithfully mirrors that of their continuum counterparts remains an open issue.

This paper is organized as follows. In Section~\ref{sec:models}, we present and analyze the dynamics of several discrete lattice models belonging to the EW and KPZ universality classes. In Section~\ref{sec:observables}, we define what is meant by a complete graph, introduce the observables considered throughout this work, and summarize the behavior obtained for these observables and the EW and KPZ equations in this case. Section~\ref{sec:results_RSOS} presents and discusses the results obtained for the RSOS model, while Section~\ref{sec:results_BD} reports the corresponding results for the BD model. Finally, concluding remarks and overall considerations are summarized in Section~\ref{sec:concl}. Some further details on simulation results and their interpretation are provided in three appendices at the end.

\section{Discrete lattice growth models}\label{sec:models}

In this section, without trying to be exhaustive, we revisit some of the most well-known discrete lattice models in the EW and KPZ classes and discuss how their dynamics simplify when they are studied on complete graphs. We begin with the EW case and then turn to several KPZ-class models, first considering those that do not lead to strictly trivial behavior on the complete-graph topology, and afterwards those whose dynamics either become trivial or cannot be naturally adapted to the topology of a complete graph.

It is worth noting that some paradigmatic models of the KPZ universality class, such as the Totally Asymmetric Simple Exclusion Process (TASEP)~\cite{Derrida1998,Spohn1991,Derrida1998,Takeuchi2018}, do not even admit a meaningful formulation on complete graphs, since their dynamics rely on the existence of preferred directions, a feature that is absent in such substrates.

\subsection{RDSR model:}

The RDSR model is one of the simplest discrete growth models belonging to the EW universality class \cite{Barabasi1995}. In this model, particles are deposited at randomly chosen sites and are allowed to relax to neighboring sites of lower height, which introduces local surface smoothing through an effective diffusive mechanism. As a result, the surface exhibits Gaussian fluctuations and behavior consistent with the EW equation \cite{Krug1997,Tauber2014}.

When this model is defined on a complete graph, it yields trivially flat surfaces. Because particles are allowed to diffuse to any neighboring site of lower height, and in a complete graph every node is adjacent to all others (see Sec.\ \ref{sec:cg} for more details on this type of substrates), no particle can be deposited at height $2$ on any node until all nodes have first reached height $1$. Consequently, as anticipated in Ref.\ \cite{Oliveira2021}, the interface grows strictly layer by layer. As a direct implication, the surface roughness does not exhibit FV scaling (in particular, its roughness resets to zero after every $N$ particle deposition events, oscillating between zero and a maximum value over each cycle). This behavior is consistent with the interpretation of the complete graph as representing the infinite-dimensional limit and from the fact that the upper critical dimension of the EW universality class is $d_u^{\mathrm{EW}} = 2$. A detailed study of the time evolution of the roughness and of the distribution of the rescaled height fluctuations for this model on a complete graph is presented in~\ref{appendix_RDSR}.

\subsection{RSOS model:}

The RSOS model is a widely studied discrete growth model belonging to the KPZ universality class \cite{Kim1989,Barabasi1995}. In this model, particles are deposited randomly onto a lattice under the constraint that the height difference between neighboring columns cannot exceed a fixed value $m$, typically chosen as one. This restriction suppresses the development of steep local slopes while preserving the nonlinear growth mechanism. As a result, the large-scale behavior of the RSOS model is described by the KPZ equation, and the interface exhibits the characteristic scaling properties and non-Gaussian fluctuations of the KPZ universality class \cite{Krug1997,Takeuchi2018}. The deposition rule of this model can be summarized as follows:
\begin{itemize}
    \item At each step, a site $i$ is selected at random from the system.
    \item The height variable $h_i$ is increased by one, provided the following condition is satisfied: 
    $$
    |h_i+1-h_j|\le m, \quad\forall j \in \mathcal{N}(i),
    $$
    where $\mathcal{N}(i)$ denotes the set of neighbors of site $i$.
    \item After each deposition attempt, time is incremented by $\Delta t = 1/N$, where $N$ is the number of sites in the substrate.
\end{itemize}
The dynamics of this system also lead to flat surfaces. In particular, for $m=1$ the evolution is very similar to that of the RDSR model, since no node can grow beyond height $2$ until all nodes have reached height $1$. This results, as in the RDSR model, in a roughness that oscillates between zero and a maximum value, although over different time intervals, since the RSOS deposition rule allows for rejection of certain deposition events. The dynamics of the RSOS model with $m=1$ are studied in detail in \ref{appendix_RSOS_m1}. Oscillations in the roughness have already been reported for the RSOS model on regular lattices in high dimensions, where it was also observed that increasing $m$ beyond $1$ smooths out these oscillations~\cite{Kim2014}.

For values of $m$ larger than $1$, the dynamics become more intricate, since the roughness is no longer constrained to return to zero. Nevertheless, it remains strictly bounded. In particular, until all nodes have reached height $1$, no node can grow beyond $1+m$, and therefore the roughness is bounded by a system-size-independent value of $m^2/4$. As a consequence, it cannot exhibit FV scaling, but at most a transient regime followed by saturation.\footnote{As will be discussed below, the oscillations described for the case $m=1$ do not disappear entirely; rather, the roughness displays small residual oscillations whose amplitude decreases as $m$ increases.} All these aspects will be discussed in detail in Section~\ref{sec:results_RSOS}.

\subsection{BD model:}

The BD model is a prototypical discrete growth model associated with the KPZ universality class \cite{Barabasi1995}. In this model, particles are released from random positions and follow vertical trajectories until they stick irreversibly upon first contact with the growing surface, producing overhangs and lateral aggregation. This aggregation mechanism generates strong correlations and a nonlinear growth process, which at large scales is described by the KPZ equation and displays the characteristic roughness scaling of the KPZ universality class \cite{Krug1997,Takeuchi2018}. On a lattice, the elementary update rules of the model are as follows:
\begin{itemize}
    \item At each step, a site $i$ is selected at random from the graph.
    \item The height variable is updated according to
    $$
    h_i = \max \{ h_i+1, h_j \}, \quad j \in \mathcal{N}(i),
    $$
    where $\mathcal{N}(i)$ denotes the set of neighbors of site $i$.
    \item After each deposition event, time is incremented by $\Delta t = 1/N$, where $N$ is the number of sites in the substrate.
\end{itemize}

Since on a complete graph all nodes $j \ne i$ are neighbors of node $i$, the simulation can be simplified by keeping track of the maximum height in the system, denoted by $h_{\mathrm{max}}$. If $h_i = h_{\mathrm{max}}$, the update rule reduces to $h_i \rightarrow h_{\mathrm{max}}+ 1$. Otherwise, the height is set to the maximum value, i.e., $h_i \rightarrow h_{\mathrm{max}}$. As we will see in Section~\ref{sec:results_BD}, the surfaces generated by this model are not trivially flat.





\subsection{PNG model:}

The Polynuclear Growth (PNG) model is a stochastic surface growth model belonging to the KPZ universality class \cite{Prhofer2000,Barabasi1995}. It describes interface evolution through random nucleation events that create new layers, which then spread laterally at a constant velocity and coalesce upon contact. This simple microscopic rule leads to highly nontrivial macroscopic behavior, including characteristic KPZ scaling exponents and Tracy--Widom distributed height fluctuations in one dimension \cite{Johansson2000,Takeuchi2018}. The dynamics of the model are defined as follows:
\begin{itemize}
  \item Random nucleation: Particles are created at randomly chosen positions on the substrate according to a Poisson process with rate $\lambda$. A nucleation event at position $x_0$ and time $t_0$ increases the local height by one unit,
  $$
    h(x_0,t_0^+) = h(x_0,t_0^-) + 1 .
  $$
  \item Deterministic lateral growth: Each nucleated island expands laterally at a constant velocity $v$, where $v$ is a model parameter. As a result, all points satisfying
  $$
    |x - x_0| \le v (t - t_0), \quad t \ge t_0,
  $$
  increase their height by one unit.
  \item Coalescence: When two expanding islands meet, they merge to form a single island, which continues to grow laterally as a whole.
\end{itemize}
On a complete graph, the behavior of this model becomes trivial. Once a nucleation event has occurred, after a short time it spreads over the entire system, since all nodes are at distance one from each other. As a consequence, this model produces completely flat interfaces.

\subsection{Etching model:}

The etching model is another discrete growth model belonging to the KPZ universality class. Originally, it was proposed to describe the dissolution of a crystalline solid by a liquid~\cite{Mello2001}. It has also been employed to mimic the erosion of a surface by an acid~\cite{Alves2016-2}. The model has two versions, one corresponding to growth and the other to erosion; however, both share the same underlying dynamics. Numerical simulations have shown that its critical exponents coincide with those of the KPZ universality class~\cite{Mello2001,Rodrigues2015,AaroReis2004}. The dynamics of the model (in its growth version) are defined as follows~\cite{Rodrigues2015}:
\begin{itemize}
    \item At each growth attempt, a site $i$ with current height $h_i \equiv h_0$ is randomly selected.
    \item The height of this site is increased by one unit, $h_i \to h_0 + 1$.
    \item Any neighboring site whose height is smaller than $h_0$ grows until its height reaches $h_0$.
\end{itemize}

When this model is implemented on a complete graph, its dynamics become trivial and lead to flat surfaces. Owing to the neighbor growth rule and to the fact that, in a complete graph, all nodes are neighbors of one another, the nodes can assume only two distinct height values: the maximum height or one unit below the maximum. In a given deposition event, two outcomes are possible: either a site that is not at the maximum height is selected, in which case its height becomes equal to the maximum, or a site that is already at the maximum height is chosen, in which case this site grows by one unit and all other sites are updated to the previous maximum height.

\section{Definitions and simulation details} \label{sec:observables}

\subsection{Complete graphs}
\label{sec:cg}

A complete graph, also known as a fully connected graph, is defined as a simple undirected network in which an edge exists between every pair of distinct vertices. As a result, all vertices have the same connectivity, with degree equal to $N-1$, where $N$ represents the total number of nodes or vertices. In contrast to tree-like topologies such as the Cayley tree, the complete graph exhibits maximal connectivity and contains the maximum number of edges allowed for a graph with $N$ vertices, namely $N(N-1)/2$. The diameter of a complete graph is equal to one, as any node can be reached from any other by traversing a single edge. Owing to this full topological equivalence of all vertices, there is no meaningful distinction between surface and bulk sites, and boundary effects are intrinsically absent. This pronounced homogeneity makes the complete graph a particularly suitable setting for assessing mean-field theories, as every node is subject to the same effective conditions dictated by the collective state of the system. As a consequence, a wide class of statistical physics models formulated on complete graphs yield results that coincide with their mean-field descriptions. Well-known examples include the Ising model, which in this limit reduces to the analytically solvable Curie--Weiss model \cite{Baxter2016} and is frequently used as a benchmark in numerical studies \cite{Newman1999}. Contact processes have also been thoroughly examined on complete graphs, where their dynamics have been shown to agree with mean-field behavior \cite{Peterson2011,Xue2017}, particularly in applications to epidemic spreading \cite{OttinoLffler2017,Guo2013}. Furthermore, numerous other models have been explored in this framework, such as voter models \cite{Lipowski2022,Azhari2022,Fronczak2017,Sood2008} and bond percolation problems \cite{Huang2018}, among many others.

\subsection{Observables}

The main observable considered in this work is the global interfacial roughness, denoted by $w(L,t)$. This quantity is defined from the second central moment of the height fluctuations according to
\begin{equation}
	\label{eq:width}
	w^2(t)=\left\langle \overline{\left[h_{i}(t)-\bar{h}(t)\right]^2} \right\rangle,
\end{equation}
where the bar $\overline{(\cdots)} \equiv (1/N)\sum_i (\cdots)$ represents an average over all spatial sites while the angular brackets $\langle \cdots \rangle$ denote an average over different realizations of the stochastic noise. 

Since every node of a complete graph is separated by a unit distance from any other, the notion of spatial separation loses significance, and spatial correlation functions cannot be properly defined. Consequently, conventional observables used in kinetic roughening, such as height–height difference correlations, reduce to quantities that are simply proportional to the interface width. Nonetheless, temporal correlations remain well defined and can be used to explore the presence of aging effects in the evolving surfaces. In particular, one can consider the two-time height autocorrelation function
\cite{Takeuchi2012,Takeuchi2018,Henkel2012},
\begin{equation}
\label{eq:ageing}
C_t(t,t_0)=\langle h_i(t)h_i(t_0)\rangle - \langle h_i(t) \rangle \langle h_i(t_0) \rangle ,
\end{equation}
where $t_0$ is an earlier (waiting) time, $t>t_0$ is the observation time and in principle there can be an implicit dependence on the particular choice of the node $i$. If all nodes are statistically equivalent, a spatial average over $i$ can be performed to improve the statistical accuracy. This observable provides direct information about memory effects~\cite{Takeuchi2012,Henkel2012,DeNardis2017} and is commonly linked to \emph{aging} behavior. Within this framework, \emph{aging} describes situations in which a system (i) relaxes slowly toward its stationary state(s) in a non-exponential fashion, (ii) does not exhibit time-translation invariance, and (iii) displays dynamical scaling~\cite{Henkel2012}. In such cases, the observable is expected to follow the scaling form
\begin{equation}
    \label{eq:escalado_aging}
    C_t(t,t_0)\sim t_0^{-b} f_C(t/t_0),
\end{equation}
where the scaling function $f_C$ is anticipated to decay asymptotically as $f_C(y)\sim y^{-\rho}$ in the  $y\rightarrow\infty$ limit \cite{Henkel2012,Rthlein2006}. The condition $C_t(t,t_0) \to 0$ as $t/t_0 \to \infty$ indicates that, although fluctuations increase over time, their temporal correlations eventually vanish, meaning that the system gradually loses memory of its initial configuration. This behavior is found for EW interfaces~\cite{Rthlein2006} and for flat KPZ interfaces~\cite{Takeuchi2012}. In contrast, circular KPZ interfaces reach a finite long-time value of the time correlation $C_t$, which reflects the existence of an \emph{infinite-time memory}; that is, trajectories remain partially correlated even asymptotically~\cite{Takeuchi2012}.

When the interface enters a stationary regime, aging effects disappear. In this situation one obtains
\begin{equation}
    C_t(t,t_0) \simeq C_{\mathrm{stat}}(t - t_0),
\end{equation}
showing that correlations depend solely on time differences, thereby restoring time-translation invariance. Furthermore, above the upper critical dimension, aging—understood as a scale-invariant dependence on the ratio $t/t_0$—is expected to be essentially absent, since correlations rapidly converge to a stationary form. This occurs in the EW equation: in one dimension correlations scale with $t/t_0$, at the critical dimension $d=2$ they display logarithmic behavior, and in dimensions above the critical one time-translation invariance is recovered~\cite{Rthlein2006}.

Beyond temporal correlations, the time power spectrum of the interface has been extensively analyzed \cite{Lauritsen1993,Takeuchi2017,VaquerodelPino2025} as a tool to infer the critical exponents characterizing the system. It is defined as \cite{Lauritsen1993}
\begin{equation}
\label{eq:power_spectra}
    S(\omega)=\frac{1}{T}\left\langle \left| \sum_{t=0}^{T-1}\bar{h}(t)e^{i\omega t} \right|^2\right\rangle.
\end{equation}
For regular substrates and dimensions below the upper critical one, where FV scaling holds, the power spectrum exhibits the scaling behavior
\begin{equation}
    S(\omega)\sim V^{-1}\omega^{-\psi},
\end{equation}
with the exponent $\psi$ given by \cite{Lauritsen1993}
\begin{equation}
\label{eq:psi}
    \psi=1+\frac{2\alpha+d}{z}.
\end{equation}

Additionally, recent progress in the field of kinetic roughening, particularly for KPZ growth \cite{Kriecherbauer2010,HalpinHealy2015,Takeuchi2018}, has shown that universality is not limited to scaling exponent values. Indeed, it has been demonstrated that suitably rescaled height fluctuations follow universal probability distributions. Introducing the normalized height variable
\begin{equation}
	\label{eq:chi}
	\chi_i(t) = \frac{h_i(t) - \bar{h}(t)}{w(t)} ,
\end{equation}
the probability density function (PDF) of $\chi$ reaches a stationary form throughout the growth regime and is shared by all models within the same universality class \cite{Kriecherbauer2010,HalpinHealy2015,Carrasco2016,Takeuchi2018,Carrasco2019}. In one spatial dimension ($d=1$), height fluctuations in the KPZ class are described by Tracy--Widom (TW) distributions, whose specific form depends e.g.\ on the interface geometry \cite{Kriecherbauer2010,HalpinHealy2015,Takeuchi2018}, while those associated with the EW class are Gaussian. For other substrates, such as complete graphs, the asymptotic distribution is still not fully understood. Above the upper critical dimension, the behavior of the height fluctuation PDF is not generally known and need not be universal. Indeed, numerical integration of the EW equation shows that the kurtosis becomes increasingly negative as the substrate dimension is raised from $d=3$ to $d=5$, whereas, over the same dimensional range, the kurtosis increases and remains positive for the discrete RDSR model, despite both systems belonging to the EW universality class and exhibiting surfaces that are essentially flat~\cite{Alves2014}. This contrasting behavior arises from the differences between the smooth continuum dynamics of the EW equation and the restricted set of accessible heights inherent to the discrete RDSR model. This comparison highlights that, above the upper critical dimension, microscopic details may remain relevant for observables like the distribution of height fluctuations and its higher-order cumulants. 
Actually, we show in \ref{appendix_RDSR} that the height fluctuations of the RDSR model on a complete graph are clearly non-Gaussian. Specifically, they are described by a perfectly symmetric bimodal distribution, with a kurtosis that diverges in the infinite-size limit. 


Statistical uncertainties for all measured observables were estimated using the jackknife resampling method~\cite{Young2015,Efron1982}. Additionally, time-dependent data were grouped into logarithmically spaced bins, within which averages were calculated. A total of 100 such time bins were typically used throughout this study. Further details regarding the jackknife procedure and the time-binning strategy are provided in Appendix B of Ref.~\cite{Barreales2020}.

\subsection{Results for the continuum equations on a complete graph}\label{sec:ec_cont}

As discussed in the Introduction, Ref.~\cite{Marcos2026} derives several analytical results for the EW equation on a complete graph. In particular, it shows that the interface roughness obeys
\begin{equation}
\label{eq:rugosidad_EW}
w^2(t)=\frac{D (N-1)}{\nu N^2}\left( 1 - e^{-2 \nu N t} \right).
\end{equation}
For short times [$t\ll\tau_\mathrm{EW}\equiv1/(2\nu N)$], the system dynamics are essentially those of the Random Deposition (RD) equation [namely, Eq.\ \eqref{eq:KPZ_orig} with $\nu=\lambda=0$], for which $w^2(t) \sim t$. In contrast, at long times the roughness saturates to a value that decreases with system size as $w^2_\mathrm{sat} \sim 1/N$.

In addition, that reference shows that the height fluctuations are Gaussian and that the time power spectrum of the spatially averaged height scales as $S(\omega)\sim \omega^{-2}$, consistent with the hyperscaling relation $2\alpha+z=d$. In particular, the scaling $S(\omega) \sim \omega^{-2}$ for the EW equation on a complete graph arises as a direct consequence of the effective Brownian dynamics of $\bar{h}(t)$ on this substrate.

Moreover, Ref.~\cite{Marcos2026} shows that the spatially averaged two-time height autocorrelation function $\overline{C_t(t,t_0)}$ admits the closed-form expression
\begin{equation}
    \label{eq:aging_pred_EW}
    \overline{C_t(t,t_0)} = w^2(t_0)\, e^{-\nu N (t-t_0)}.
\end{equation}
This set of results not only provides an accurate characterization of the EW equation on a fully connected graph, but it also yields an excellent approximation to the KPZ equation on the same graph in the large-system-size limit \cite{Marcos2026}.

\section{Simulation Results for the RSOS Model} \label{sec:results_RSOS}

In this section, we discuss the results of the simulations performed for the RSOS model with $m\ne1$, which happens not to be trivial on fully-connected graphs.

\subsection{Roughness and height fluctuations}

\begin{figure}[!t]
\centering
\includegraphics[width=0.6\textwidth]{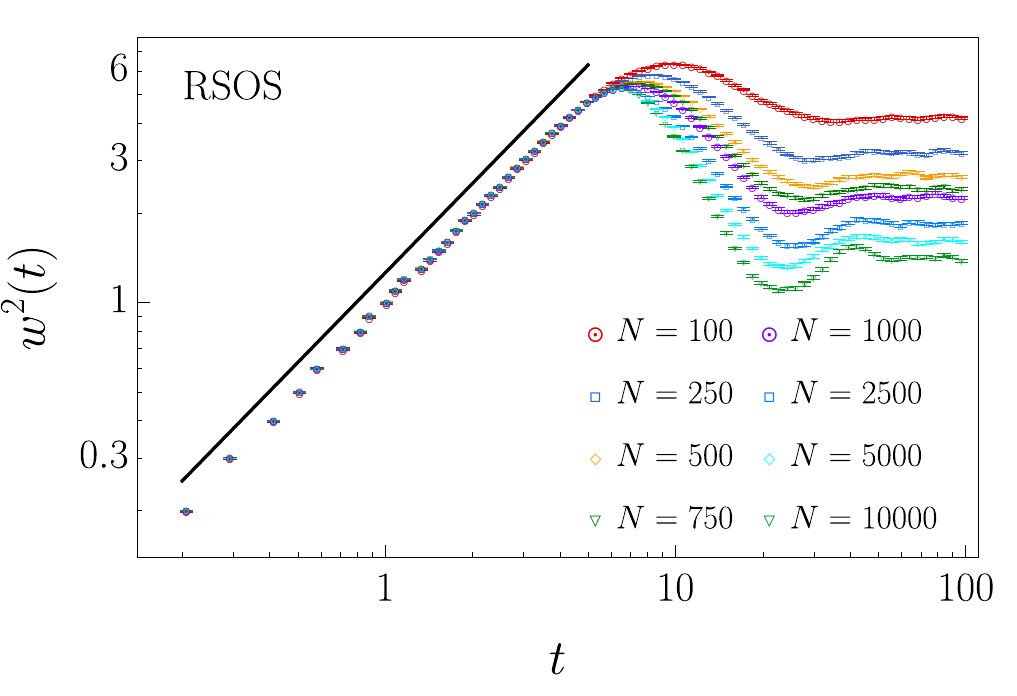}
\caption{Log--log plot of the squared roughness, $w^2(t)$, as a function of $t$ for the RSOS model. Data are shown for different system sizes $N$ (see legend) and $m=10$. The solid black line corresponds to the RD scaling $w^2(t) \sim t$.}
\label{fig:w_RSOS}
\end{figure}

Figure~\ref{fig:w_RSOS} shows the time evolution of the squared roughness, $w^2(t)$, for the RSOS model at different system sizes and $m=10$. For short times ($t\ll m$), the squared roughness grows as in RD, i.e., $w^2(t) \sim t$. However, the behavior at longer times is more complex to analyze. First, a pronounced rebound is observed, during which the squared roughness rapidly decreases before stabilizing around an approximately time-independent value.
Similar non-monotonic behavior with time of the surface roughness is well documented for epitaxial surface growth \cite{Kallabis1997,Krug1997,Michely2004} and has also been assessed, e.g., in the context of model A critical dynamics for the 2D Ising universality class \cite{VaquerodelPino2025}. In our present context, this behavior can be expected upon inspection of the microscopic system dynamics. Indeed, the deposition rule
\begin{equation}
   |h_i+1-h_j|\le m, \quad\forall j \in \mathcal{N}(i),
\end{equation}
is readily seen to be equivalent to a much simpler one. After a deposition attempt at site $i$, it is only necessary to check the nodes $j$ with smaller height, since the condition with nodes of larger height was already satisfied before the attempt. Among the nodes with smaller height, it suffices to compare with the global minimum height $h_\mathrm{min}$. The condition then reduces to
\begin{equation}
   h_i+1\le h_\mathrm{min}+m \qquad \Leftrightarrow \qquad h_i-h_\mathrm{min}<m.
\end{equation}
In other words, it is enough to verify whether the difference between the height of node $i$ and the global minimum is smaller than $m$ (in which case the deposition is accepted) or larger than or equal to $m$ (in which case it is rejected).

For short times, when most nodes have not yet reached height $m$, this condition is always satisfied. Consequently, all deposition attempts are accepted and the growth proceeds as in the RD model; therefore, the roughness evolves as $w^2(t) \sim t$. Only once a significant fraction of nodes has reached height $m$, and deposition events at those nodes begin to be rejected, does the roughness start to deviate from this behavior. Shortly thereafter, when many nodes have already reached height $m$ but some still remain at their initial height, the roughness attains its maximum value. From that point onward, deposition events are accepted only at nodes that lag significantly behind the rest, and their subsequent growth leads to a decrease in the overall roughness.

The end of the RD-like transient is mainly controlled by the value of $m$, although it also depends more weakly on $N$. The larger the system size, the more likely some nodes reach the height ceiling set by $m$ and begin to slow down the overall growth. Approximately, one can say that in the RSOS model the characteristic timescale controlling the dynamics is $\tau_\mathrm{RSOS}\lesssim m$.

The decrease in roughness eventually comes to an end at somewhat longer times, when all nodes have left their initial height and those previously blocked at height $m$ can start to grow again. 

At long times, the roughness continues to exhibit oscillatory behavior. For a given noise realization, it fluctuates between its minimum value ($0$) and its maximum value ($m^2/4$). When the condition $h_{\mathrm{max}} - h_{\mathrm{min}} = m$ is satisfied, growth at nodes with maximum height becomes blocked, forcing the roughness to decrease. When this condition is not satisfied, growth at these nodes is allowed again, leading to an increase in the roughness. However, when averaged over many noise realizations, the roughness approaches a stationary value, although remnants of these oscillations remain visible in the averaged quantity.

\begin{figure}[!t]
\centering
\includegraphics[width=0.6\textwidth]{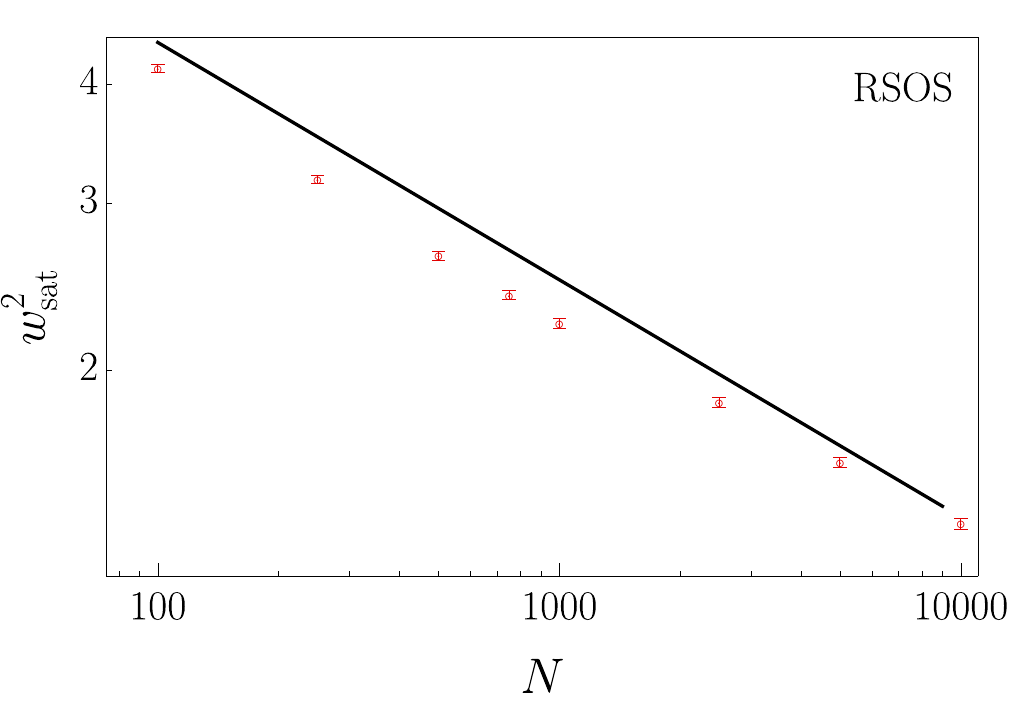}
\caption{Log--log plot of the saturation value of the square of the roughness, $w^2_\mathrm{sat}$, as a function of the system size $N$ for the RSOS model. In this plot, $m=10$. The solid black line corresponds to $w^2_\mathrm{sat} \sim N^{-1/4}$ scaling.}
\label{fig:w_RSOS_sat}
\end{figure}

Figure~\ref{fig:w_RSOS_sat} shows the saturation value of the squared roughness, $w^2_\mathrm{sat}$, as a function of the system size $N$. The reference line shown in this figure, corresponding to $w^2_\mathrm{sat}\sim N^{-1/4}$, is included only as a guide to the eye to illustrate the overall trend of the data, rather than as evidence of a definite asymptotic scaling law. Indeed, the data for larger system sizes display a certain curvature in this representation, indicating that the finite-size dependence of $w^2_\mathrm{sat}$ is not yet fully resolved within the range of system sizes accessible in our simulations. In particular, a slower decay, possibly of logarithmic form, cannot be ruled out. This possibility is consistent with the exact result for the case $m=1$ [see Eq~\eqref{eq:w_sat_RSOS_m1}]. However, for $m\neq 1$ the present numerical data do not allow us to unambiguously determine the asymptotic behavior, and therefore we prefer not to advocate any specific functional form.

\begin{figure}[!t]
\centering
\includegraphics[width=0.6\textwidth]{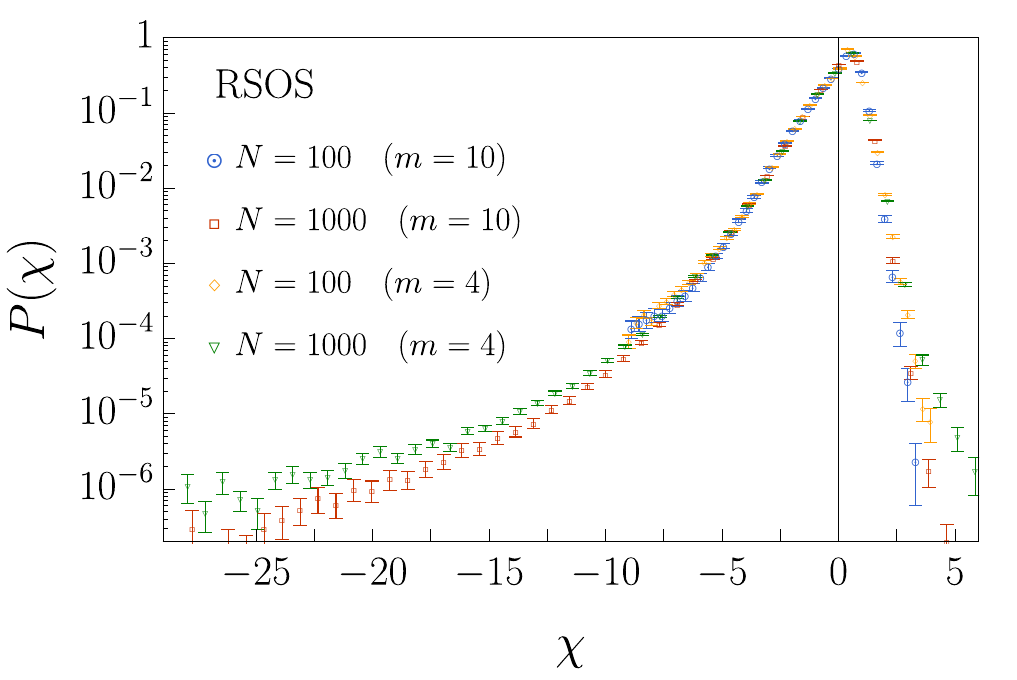}
\caption{Histogram of the rescaled height fluctuations $\chi$ [see Eq.~\eqref{eq:chi}], for the RSOS model. Blue circles and red squares correspond to $m=10$ with $N=100$ and $N=1000$, respectively, whereas orange diamonds and green triangles correspond to $m=4$ with $N=100$ and $N=1000$, respectively.}
\label{fig:chi_RSOS}
\end{figure}

Turning to the analysis of height fluctuations, we find (not shown) that during the RD growth regime the $h_i$ are distributed according to
\begin{equation}
	h_i \sim \mathrm{Poisson}(t),
\end{equation}
which corresponds to the expected behavior for the RD model \cite{Barabasi1995}. In stark contrast, at the final stationary state, Fig.~\ref{fig:chi_RSOS} shows the distribution of the height fluctuations [see Eq.~\eqref{eq:chi}] for the RSOS model for two values of the parameter, $m=10$ and $m=4$, and two system sizes, $N=100$ and $N=1000$. These fluctuations exhibit a markedly extended left tail, while the right tail decays much more rapidly. This behavior provides significant insight into the system dynamics in this long-time regime. In particular, it indicates that in this state the mean height $\bar{h}$ lies very close to the maximum height. Therefore, many nodes remain significantly lagging in their growth while the maximum height stays saturated. Specifically, the nodes at the global minimum constrain the system dynamics. If there are $k$ such nodes, the typical time for one of them to leave the global minimum is $\tau_k=1/k$. As a consequence, these lagging states persist for increasingly long times and effectively freeze the dynamics of the entire system, giving rise to the extended left tail of the distribution. Until all nodes have escaped from the global minimum, the global maximum cannot increase, which leads to an accumulation of nodes around this maximum value.

As shown in Fig.~\ref{fig:chi_RSOS}, the rescaled height-fluctuation distributions for the two system sizes overlap reasonably well over most of the accessible range, indicating only a weak dependence on $N$. The main effect of increasing the system size is to improve the sampling of rare fluctuations, thereby allowing the tails of the distribution to be explored over a broader range. The distributions corresponding to $m=4$ and $m=10$ also exhibit a similar overall shape, although some dependence on $m$ remains visible in both tails, where fluctuations farther from the mean appear to be slightly more probable for the smaller value of $m$.


\subsection{Time power spectra and aging properties}

\begin{figure}[!t]
\centering
\includegraphics[width=0.6\textwidth]{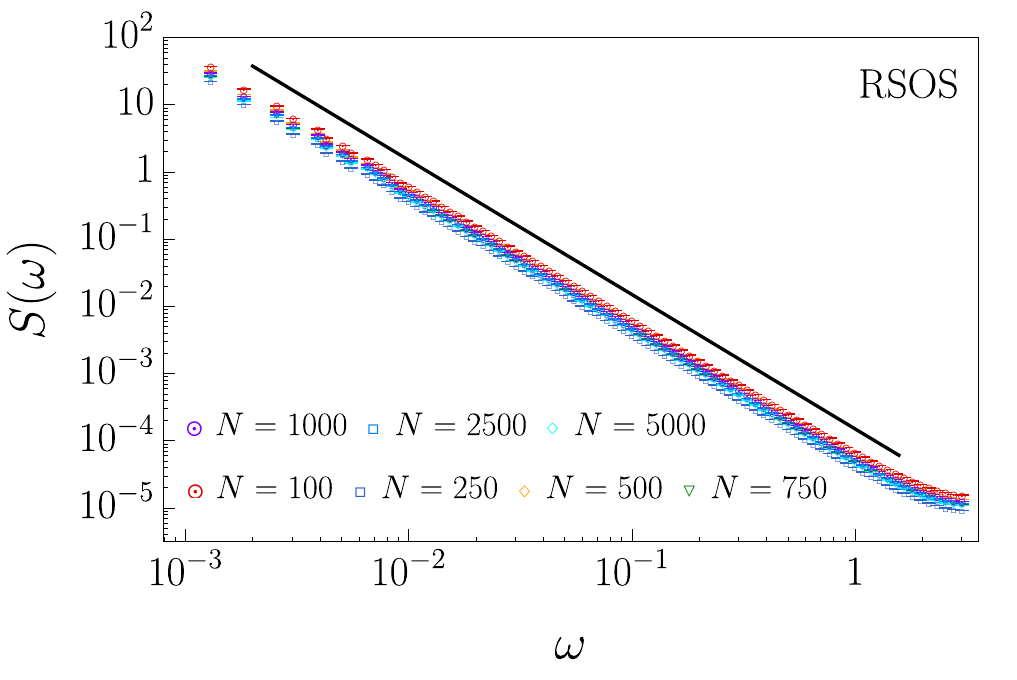}
\caption{Time power spectrum $S(\omega)$ [see Eq.~\eqref{eq:power_spectra}] for the RSOS model. Data are shown for different system sizes $N$ (see legend) and $m=10$. The solid black line corresponds to $S(\omega)\sim\omega^{-2}$ scaling.}
\label{fig:PS_RSOS}
\end{figure}

Figure~\ref{fig:PS_RSOS} shows the time power spectrum, $S(\omega)$, for several system sizes of the RSOS model. The slope observed in the figure follows $S(\omega) \sim \omega^{-2}$ quite closely, which coincides with the behavior observed for the EW and KPZ equations on a complete graph, as explained in Sec.~\ref{sec:ec_cont}. This result is consistent with the hyperscaling relation $2\alpha + z = d$, which is characteristic of linear equations, and contrasts with the Galilean invariance relation $\alpha + z = 2$ associated with the KPZ fixed point.

\begin{figure}[!t]
\centering
\includegraphics[width=0.6\textwidth]{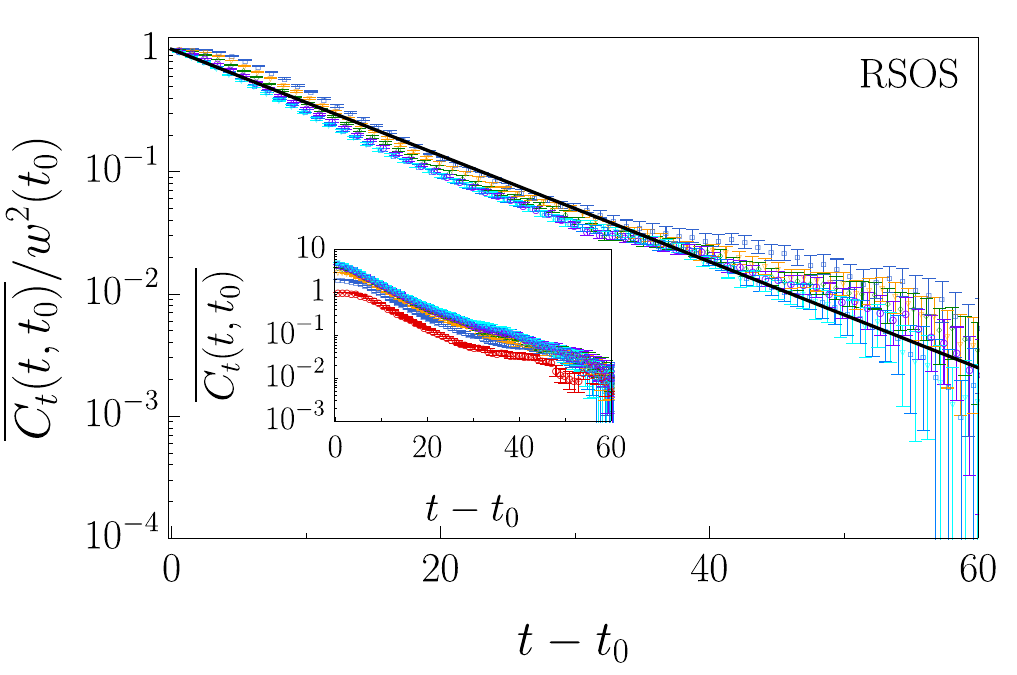}
\caption{Spatially averaged two-time height autocorrelation function, rescaled by the squared roughness, $\overline{C_t(t,t_0)}/w^2(t_0)$, as a function of the time difference $t-t_0$ for the RSOS model, shown for different waiting times $t_0=\{1, 2, 3, 4, 5, 6, 7\}$, which appear bottom to top in the inset. The system size is $N=1000$ and $m=10$. The solid black line corresponds to the theoretical prediction given by Eq.~\eqref{eq:aging_RSOS} with $\tau_\mathrm{RSOS}=m=10$. Inset: Same data shown without rescaling by the squared roughness $w^2(t_0)$.}
\label{fig:ageing_RSOS_t_cortos}
\end{figure}

Figure~\ref{fig:ageing_RSOS_t_cortos} shows the spatially averaged two-time height autocorrelation function rescaled by the squared roughness,
$\overline{C_t(t,t_0)}/w^2(t_0)$, as a function of the time difference $t-t_0$, for a fixed system size of the RSOS model and several waiting times $t_0$ smaller than the typical relaxation time of the RSOS model, $\tau_\mathrm{RSOS}\lesssim m$. The behavior expected for this observable is given by Eq.~\eqref{eq:aging_pred_EW}, in which the spatially averaged two-time height autocorrelation function, rescaled by the squared roughness, should decay exponentially as a function of the time difference $t-t_0$. The prefactor in the exponential is $\nu N$, which corresponds to the characteristic timescale controlling the dynamics of the EW and KPZ equations on large complete graphs. For the RSOS model, indeed the behavior of the numerical data seems to be adequately described by
\begin{equation}\label{eq:aging_RSOS}
    \overline{C_t(t,t_0)}/w^2(t_0)=e^{-(t-t_0)/\tau_\mathrm{RSOS}}.
\end{equation}
This corresponds to the solid black line shown in Fig.~\ref{fig:ageing_RSOS_t_cortos}, where $\tau_\mathrm{RSOS}=m=10$ has been used. It is important to note that this observable exhibits significant oscillations, which makes the data collapse less clean than that observed for the continuum equations.

\begin{figure}[!t]
\centering
\includegraphics[width=0.6\textwidth]{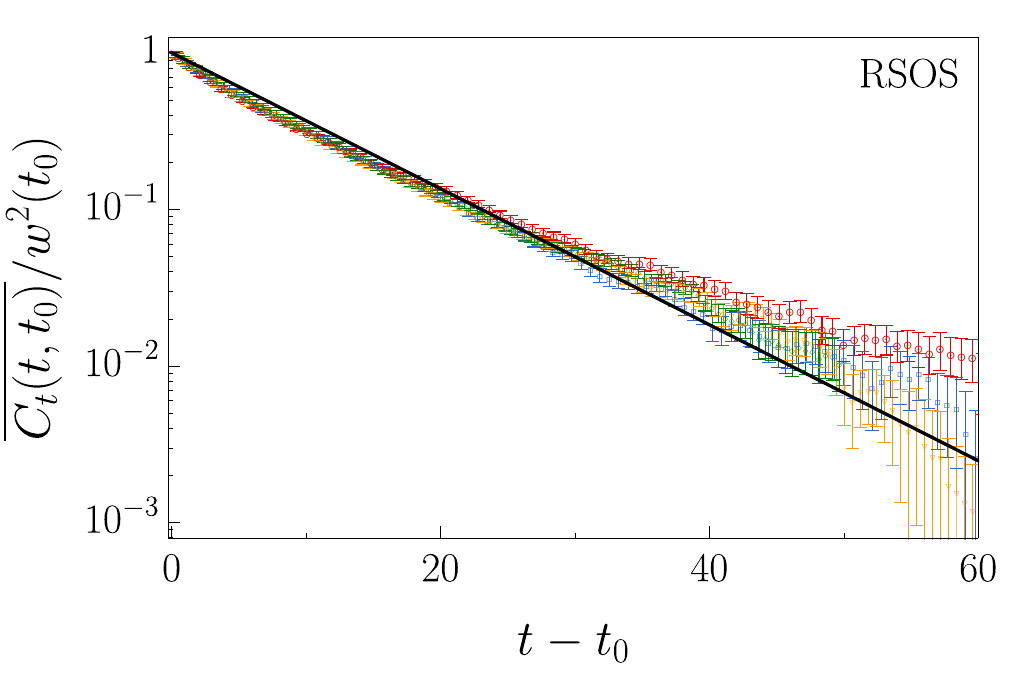}
\caption{Spatially averaged two-time height autocorrelation function, rescaled by the squared roughness, $\overline{C_t(t,t_0)}/w^2(t_0)$, as a function of the time difference $t-t_0$ for the RSOS model, shown for different waiting times $t_0=\{20,30,40,50\}$. The system size is $N=1000$ and $m=10$. The solid black line corresponds to the theoretical prediction given by Eq.~\eqref{eq:aging_RSOS} with $\tau_\mathrm{RSOS}=m=10$.}
\label{fig:ageing_RSOS_t_largos}
\end{figure}

For waiting times longer than the typical relaxation time $\tau_\mathrm{RSOS}$, the behavior remains the same. However, since the roughness has already saturated to an approximately constant value at those times, the curves collapse whether they are rescaled by the roughness or not. Figure~\ref{fig:ageing_RSOS_t_largos} shows the spatially averaged two-time height autocorrelation function rescaled by the squared roughness,
$\overline{C_t(t,t_0)}/w^2(t_0)$, for these large waiting times.

The interpretation of aging in the RSOS model is analogous to that found for the EW and KPZ continuum equations and stems from the presence of a single relaxation timescale controlling the fluctuations. Temporal correlations quickly become effectively time-translation invariant, $C_t(t,t_0) \simeq C_{\mathrm{stat}}(t - t_0)$, and, for fixed $t_0$, the two-time autocorrelation function vanishes as $t - t_0 \to \infty$, meaning that trajectories decorrelate over a finite timescale. Furthermore, no nontrivial aging collapse is observed in terms of the scaling variable $t/t_0$. In this sense, similarly to the continuum equations, the RSOS model on a complete graph exhibits at most \emph{trivial aging} during the short pre-asymptotic regime in which the roughness has not yet reached saturation—reflected in the $t_0$ dependence of the prefactor $w^2(t_0)$—while the scale-free aging typical of low-dimensional substrates is essentially absent.

\subsection{Discussion}

A summary of the results for this model is as follows. The behavior of the system is controlled by a single relaxation time, $\tau_\mathrm{RSOS}$. For times shorter than this scale, the dynamics are essentially the same as in the RD model, characterized by a growth of the roughness $w(t)\sim t^{1/2}$ and height fluctuations distributed as $h_i\sim\mathrm{Poisson}(t)$. This relaxation time $\tau_\mathrm{RSOS}$ is somewhat smaller than the value $m$ that limits the growth of nodes in the system. 

At later times the system reaches saturation at a value that decreases with the system size, approximately as $w^2_\mathrm{sat}\sim N^{-1/4}$. In this regime, the height fluctuations display a pronounced left tail as a consequence of nodes that lag behind in the growth process, effectively freezing the dynamics. This behavior is non-universal and depends weakly on both $N$ and $m$.

Surprisingly, the time power spectrum follows the same scaling as that found for the continuum equations, $S(\omega)\sim \omega^{-2}$, with particular accuracy. Moreover, the aging behavior of the system is essentially trivial, since temporal correlations quickly become effectively time-translation invariant, with an exponential decay characterized by the typical relaxation time of the model, $\tau_\mathrm{RSOS}$.

In general, the results for this model are similar to those obtained for the continuum EW and KPZ equations on a large complete graph: RD-like growth at short times, saturation values that decrease with system size, a time power spectrum scaling as $S(\omega)\sim \omega^{-2}$, and essentially trivial aging. However, there remain important differences between the continuum equations and the discrete model, particularly in the saturation regime. In the case of the former, the stationary state is characterized by an efficient suppression of fluctuations in the system by the Laplacian term (surface tension). This leads to very flat interfaces across different noise realizations and to Gaussian rescaled height fluctuations. By contrast, in the RSOS model the long-time behavior is characterized by alternating periods of time within which the growth of the maximum height is hampered, forcing the roughness to decrease, and periods within which this growth is re-enabled, causing the roughness to increase. As a consequence, the final state of the RSOS model is qualitatively different from that exhibited by the continuum equations. While the roughness averaged over many noise realizations approaches a stationary value, individual realizations display large oscillations. In addition, the rescaled height fluctuations exhibit a pronounced, non-Gaussian left tail.

\section{Simulation Results for the BD Model} \label{sec:results_BD}

In this section, we discuss the results of the simulations performed for the BD model which, as in the RSOS case, remains non-trivial on fully-connected graphs.

\subsection{Roughness and height fluctuations}

\begin{figure}[!t]
\centering
\includegraphics[width=0.6\textwidth]{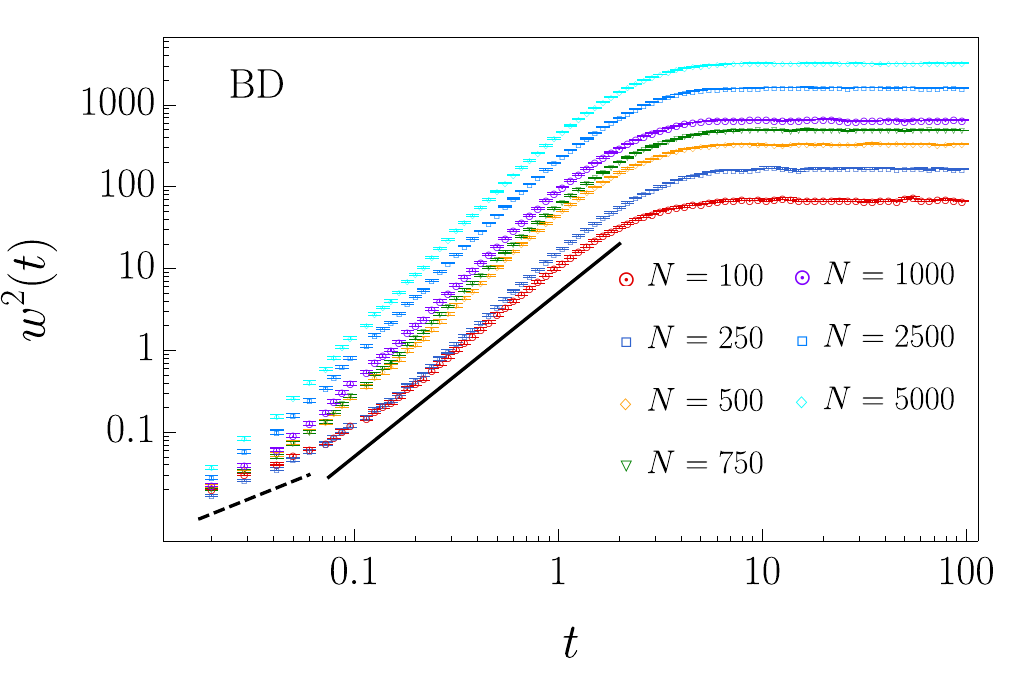}
\caption{Log--log plot of the square of the roughness, $w^2(t)$, as a function of $t$ for the BD model. Data are shown for different system sizes $N$ (see legend). The solid black line corresponds to $w^2(t) \sim t^2$ scaling, while the dashed black line corresponds to $w^2(t) \sim t$ behavior.}
\label{fig:w_BD}
\end{figure}

\begin{figure}[!t]
\centering
\includegraphics[width=0.6\textwidth]{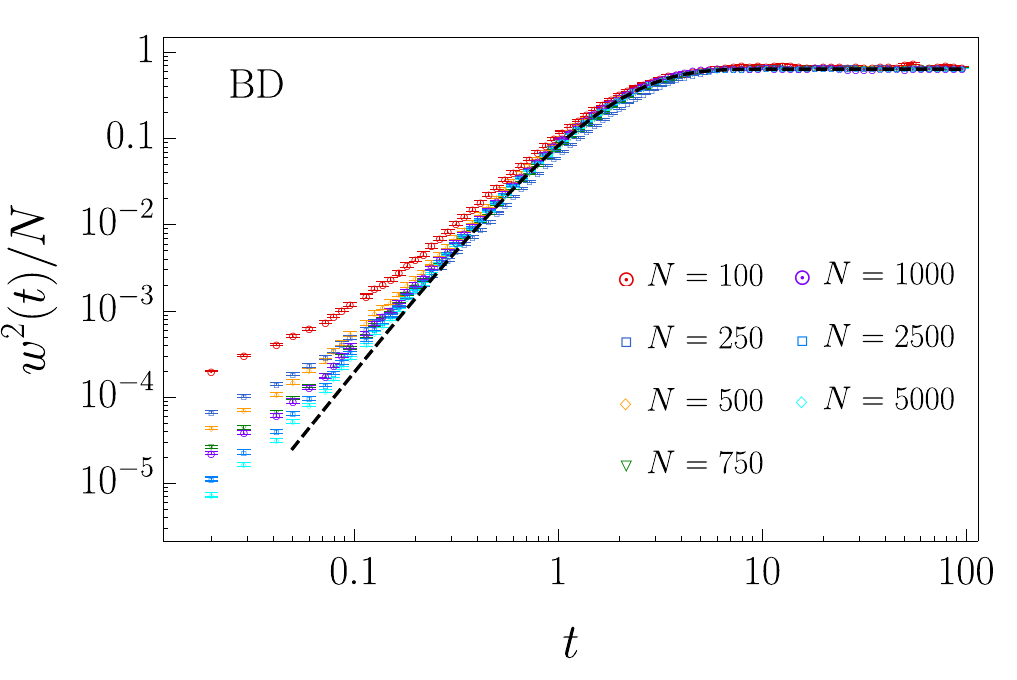}
\caption{Log--log plot of the square of the roughness rescaled by the system size, $w^2(t)/N$, as a function of $t$ for the BD model. Data are shown for different system sizes $N$ (see legend). The dashed black line corresponds to the theoretical prediction given by Eq.~\eqref{eq:rugosidad_BD}.}
\label{fig:w_BD_colapso_2}
\end{figure}

Figure~\ref{fig:w_BD} shows the time evolution of the squared global roughness, $w^2(t)$, for the BD model and different system sizes, while Fig.~\ref{fig:w_BD_colapso_2} shows the time evolution of the squared global roughness normalized by the system size, $w^2(t)/N$. In Fig.~\ref{fig:w_BD}, two clearly different growth regimes are observed. First, there is a very short RD-like transient, indicated by the dashed line, characterized by $w^2(t)\sim t$, which becomes increasingly difficult to observe as the system size increases. This regime is followed by an ultrafast growth stage, indicated by the solid line, in which the squared global roughness grows approximately as $w^2(t)\sim t^2$. Finally, at longer times, the squared global roughness reaches a time-independent value proportional to the system size, $w^2_{\mathrm{sat}}\sim N$, as clearly shown in Fig.~\ref{fig:w_BD_colapso_2}.

From these two figures, one might conclude that the critical exponents of this system are $\beta = 1$, $\alpha = 1/2$, and $z = 0$, which is contradictory and clearly inconsistent with the FV scaling behavior expected for the KPZ universality class, for which $\beta=1/3$ in one dimension and decreases monotonically with dimension \cite{Oliveira2022}. Additionally, this set of exponent values does not fulfill the Galilean scaling relation $\alpha + z = 2$. Yet, $z = 0$ seems natural on a complete graph, as all sites are at unit distance and therefore there are no lateral correlations to spread across the system.

However, the ultrafast behavior of the roughness prior to steady state does not constitute a genuine growth regime either. This is evident from the fact that it is entirely confined to the submonolayer regime, $t\lesssim 2$, where the BD model on regular lattices is already known to exhibit similar transients, characterized by faster growth than that of RD, before entering a genuine dynamic-scaling regime \cite{Banerjee2014,Mal2016}. Moreover, as shown analytically in \ref{appendix_BD_detail}, such a non-standard growth behavior originates from a transient in which deposition events cause the maximum height of the system to increase rapidly while many nodes remain trapped at $h_i=0$. Once most nodes have received a few deposition events, the roughness is forced to become time-independent. More precisely, in this intermediate ultrafast-growth regime, the squared roughness behaves as
\begin{equation}
\label{eq:rugosidad_BD}
w^2(t)=\frac{2N}{\pi}\left(1-e^{-2t}-2te^{-t}\right),
\end{equation}
which corresponds exactly to the dashed line shown in Fig.~\ref{fig:w_BD_colapso_2}, see \ref{appendix_BD_detail} for a detailed derivation.

It is important to emphasize that the ultrafast transient growth observed in the BD model has a geometric, rather than a dynamical-critical, origin~\cite{Banerjee2014,Mal2016}. At very early times, $t\ll 1$, the surface has received only a small number of particles, and the columns are not yet correlated. As deposition proceeds, the first columns with the overhang structures characteristic of the BD model begin to form. In this regime, each additional particle that attaches near a high point can make a disproportionately large contribution to the roughness, since only a few layers have been deposited and the height statistics are still determined by only a few deposition events. This is therefore a finite-time effect, rather than a universal KPZ growth exponent, and it disappears once a sufficiently large number of deposition events has occurred.



\begin{figure}[!t]
\centering
\includegraphics[width=0.6\textwidth]{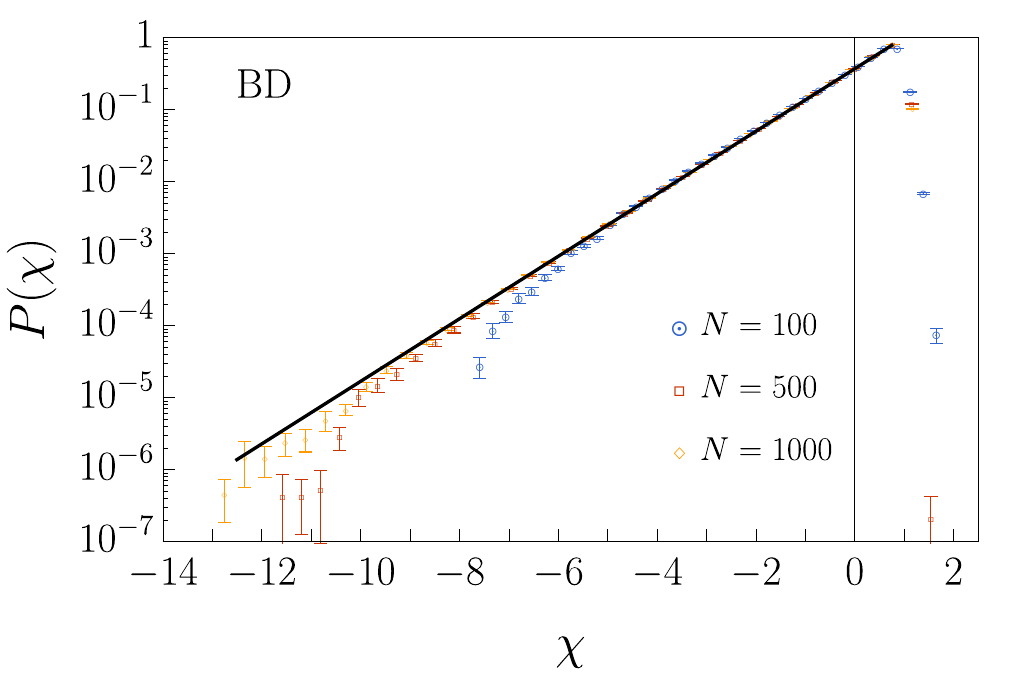}
\caption{Histogram of the rescaled height fluctuations $\chi$ [see Eq.~\eqref{eq:chi}] for the BD model. Blue circles, red squares, and orange diamonds correspond to $N=100$, $N=500$, and $N=1000$, respectively. The solid black line corresponds exactly to the distribution $P(\chi)=e^{\chi-1}$.}
\label{fig:chi_BD}
\end{figure}

Figure~\ref{fig:chi_BD} presents the height fluctuations in the stationary regime, i.e., once the roughness has become time-independent, for the BD model and for three different system sizes, namely $N=100$, $N=500$, and $N=1000$. The height fluctuations also display a markedly non-standard behavior, characterized by a pronounced exponential tail for negative values of $\chi$. As shown in detail in \ref{appendix_BD_detail}, the corresponding distribution is predicted to be
\begin{equation}
P(\chi)=e^{\chi-1},\qquad\chi<1.
\end{equation}
This feature departs significantly from the distributions usually expected in this context, where Gaussian or similar forms are typically anticipated, and indicates (similarly to what occurs in the RSOS model) the presence of nodes that remain significantly lagging behind in their growth, while others accumulate around the maximum height. However, due to the sticking rule of the BD model, this phenomenon is even more pronounced. Moreover, this behavior appears to be largely independent of the system size. As shown in Fig.~\ref{fig:chi_BD}, the distributions obtained for different values of $N$ are consistent with the same underlying exponential behavior. The main effect of increasing the system size is to improve the sampling of rare events, thereby allowing the exponential tail to be probed further into the region of very small probabilities.

A detailed analysis of the microscopic behavior of the model is presented in \ref{appendix_BD_detail}, where we examine the dynamics and temporal growth of the maximum height, the evolution of the roughness from the initial RD transient through the ultrafast-growth regime to its stationary value, the origin of the exponential left tail of the rescaled height-fluctuation distribution, and the typical relaxation time of the system, for which we also provide a simple estimate.

\subsection{Time power spectra and aging properties}

\begin{figure}[!t]
\centering
\includegraphics[width=0.6\textwidth]{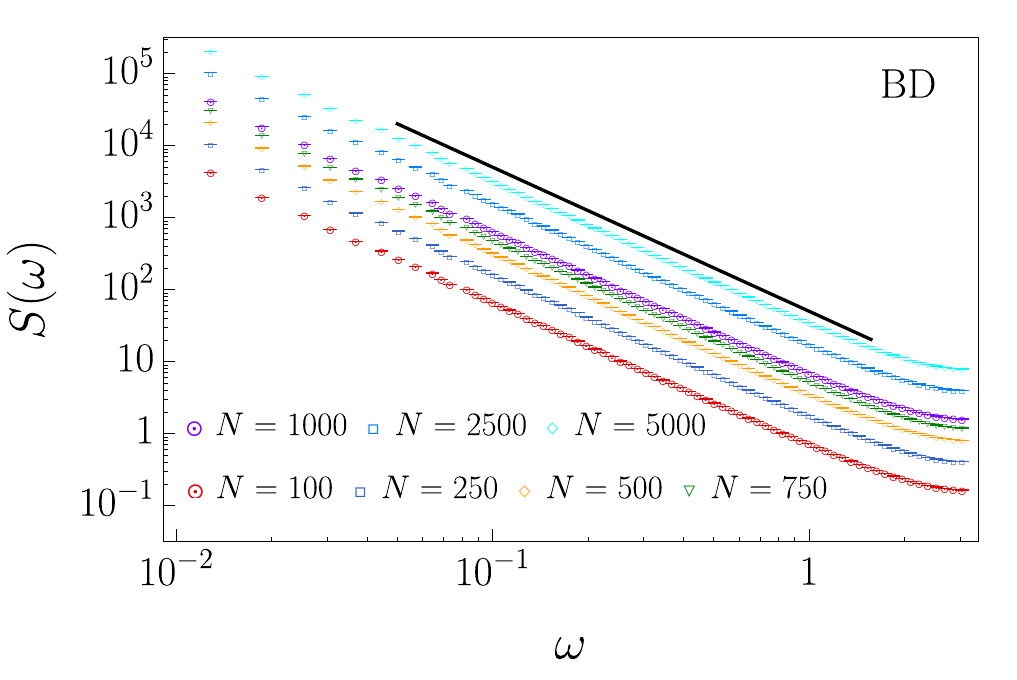}
\caption{Time power spectrum $S(\omega)$ [see Eq.~\eqref{eq:power_spectra}] for the BD model. Data are shown for different system sizes $N$ (see legend). The solid black line corresponds to  $S(\omega)\sim\omega^{-2}$ scaling.}
\label{fig:PS_BD}
\end{figure}

Figure~\ref{fig:PS_BD} shows the time power spectra, $S(\omega)$ of the BD model for several system sizes. The slope observed in the figure follows $S(\omega) \sim \omega^{-2}$, which coincides with the behavior observed for the EW and KPZ equations on a complete graph as explained in Sec.~\ref{sec:ec_cont}. As said for the RSOS model, this result is consistent with the hyperscaling relation $2\alpha + z = d$, which is characteristic of linear equations, and contrasts with the Galilean invariance relation $\alpha + z = 2$ associated with the KPZ fixed point.

\begin{figure}[!t]
\centering
\includegraphics[width=0.6\textwidth]{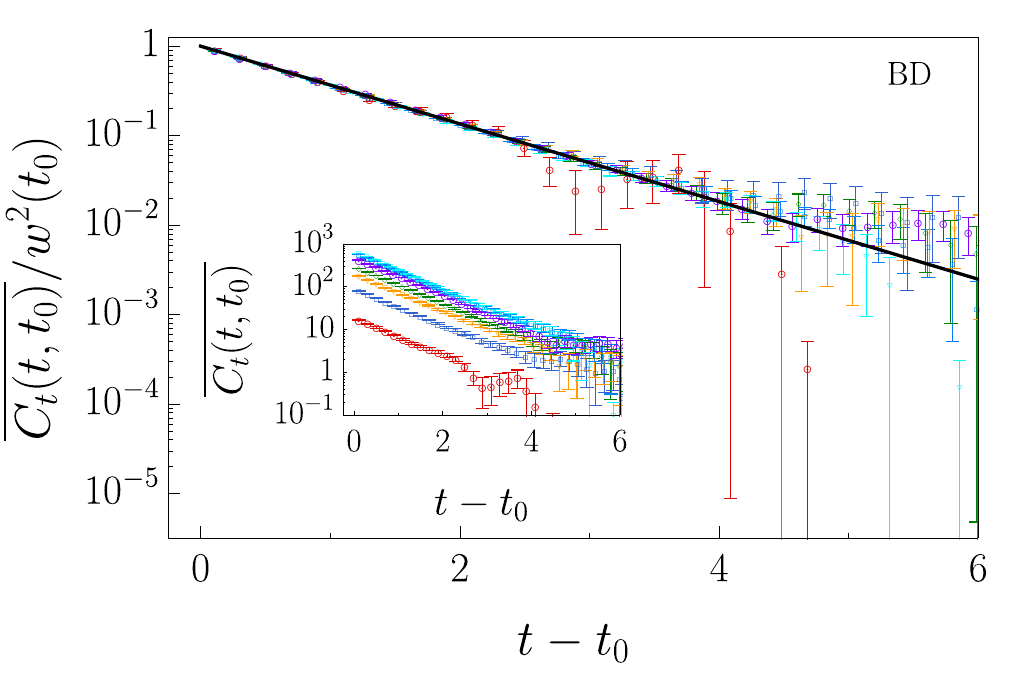}
\caption{Spatially averaged two-time height autocorrelation function, rescaled by the squared roughness, $\overline{C_t(t,t_0)}/w^2(t_0)$, as a function of the time difference $t-t_0$ for the BD model, shown for different waiting times $t_0=\{0.5,1, 1.5, 2, 3,5 ,10\}$, which appear bottom to top in the inset. The system size is $N=1000$. The solid black line corresponds to the theoretical prediction given by Eq.~\eqref{eq:aging_BD}. Inset: Same data shown without rescaling by the squared roughness $w^2(t_0)$.}
\label{fig:ageing_BD}
\end{figure}

Figure~\ref{fig:ageing_BD} shows the spatially averaged two-time height autocorrelation function rescaled by the squared roughness,
$\overline{C_t(t,t_0)}/w^2(t_0)$, as a function of the time difference $t-t_0$, for a fixed system size of the BD model and several waiting times $t_0$.

The behavior expected for this observable is given by Eq.~\eqref{eq:aging_pred_EW}, according to which the spatially averaged two-time height autocorrelation function, rescaled by the squared roughness, should decay exponentially as a function of the time difference $t-t_0$. In that expression, the prefactor in the exponential is $\nu N$, which sets the characteristic timescale governing the dynamics of the continuum equations. For the BD model, we find that the behavior 
\begin{equation}\label{eq:aging_BD}
    \overline{C_t(t,t_0)}/w^2(t_0)=e^{-(t-t_0)},
\end{equation}
which corresponds to the solid black line shown in Fig.~\ref{fig:ageing_BD}, does provide an adequate description of the numerical data. This result suggests that the characteristic relaxation time associated with this observable in the BD model is $\tau_\mathrm{BD}=1$.

In this context, the interpretation of aging in this model parallels that of the EW and KPZ continuum equations and the RSOS model and originates from the existence of a single relaxation scale governing the fluctuations. The fluctuations relax toward stationarity at a characteristic rate (equal to $\nu N$ in the continuum description and to unity in the BD model). As a result, temporal correlations rapidly become effectively time-translation invariant, $C_t(t,t_0) \simeq C_{\mathrm{stat}}(t - t_0)$, and for fixed $t_0$ the two-time autocorrelation function decays to zero in the $t - t_0 \to \infty$ limit; that is, trajectories decorrelate over a finite timescale. Moreover, no nontrivial aging collapse is observed in terms of the scaling variable $t/t_0$. In this sense, as in the case of the continuum equations, the complete-graph BD model displays, at most, \emph{trivial aging} during the short pre-asymptotic regime in which the roughness has not yet reached its stationary value---manifested through the $t_0$-dependence of the prefactor $w^2(t_0)$---, whereas the scale-free aging characteristic of low-dimensional substrates is essentially absent.

\subsection{Discussion}

A summary of the results for the BD model is as follows. After a very short RD-like transient (dashed line in Figure~\ref{fig:w_BD}), the dynamics are dominated by events in which the system maximum height increases, leaving many nodes behind that receive no particles and thus remain at low height values, together with events in which such delayed nodes jump directly to the current maximum height in a single deposition event (see \ref{appendix_BD_detail} for details). This ultrafast-growth regime, observed over the $0.1 \lesssim t \lesssim 2$ interval, is characterized by a much more rapid increase in the squared roughness $w^2(t)$ that is much faster than in RD, without a clearly defined scaling exponent, as described by Eq.~\eqref{eq:rugosidad_BD}. It rapidly disappears and gives way to a time-independent roughness after just a few deposition events per node. Moreover, the fact that many nodes are left behind and later reach the maximum height in a single step leads to the emergence of an exponential tail in the distribution of the height fluctuations. This behavior differs markedly from that obtained for the EW and KPZ equations on a complete graph, where the roughness rapidly saturates after an exponential transient with characteristic time $(\nu N)^{-1}$, and the saturation value scales as $1/N$, while the fluctuations become Gaussian in the large-size limit.

On the other hand, the time power spectrum exhibits a behavior similar to that of the EW and KPZ equations when studied on a complete graph, namely $S(\omega) \sim \omega^{-2}$. Finally, the two-time height autocorrelation function follows the same behavior as in the EW and KPZ equations, as correlations rapidly become effectively stationary and the scale-free aging characteristic of low-dimensional substrates is absent.

In general, the results for this model differ markedly from those obtained for the continuum equations on a complete graph. Beyond the RD-like growth observed at short times, the roughness exhibits an ultrafast-growth regime reminiscent of the anomalously rapid growth transients known to occur in the BD model on regular lattices, and reaches time-independent values that do not decrease with system size but instead increase with $N$. In addition, the rescaled height fluctuations display an exponential left tail, both at long times and during the ultrafast growth transient. The only similarities with the continuum equations are the time power spectrum scaling as $S(\omega)\sim \omega^{-2}$ and the essentially trivial aging behavior.


The unusual dynamics observed in this model arises from extending the BD sticking rule to a complete-graph substrate. The ultrafast-growth regime is reminiscent of the short-time transients observed for the BD model on regular substrates in finite dimensions and should therefore not be regarded as a new scaling regime specific to the complete graph. In particular, this transient behavior does not imply that the BD model falls outside the KPZ universality class in high dimensions. At later times, $t\gg1$, when a genuine growth regime would normally be expected, the roughness rapidly approaches a time-independent value. The resulting absence of a conventional FV growth regime is instead consistent with the effectively infinite-dimensional character of the complete-graph substrate.

It is worth noting that a nontrivial dependence on substrate dimensionality is not unusual in nonequilibrium growth models. For instance, the Wolf--Villain and Das Sarma--Tamborenea models are known to exhibit changes in their large-scale behavior as the substrate dimension increases, and analogous dimensional effects have also been reported for continuum growth equations~\cite{DasSarma2002,Nicoli2013}. The connection between the BD model and the KPZ equation is likewise known to involve substantial finite-time and finite-size corrections in higher dimensions, as well as technical difficulties in deriving a continuum description directly from the microscopic rules~\cite{Alves2016,Katzav2004,Haselwandter2006}.

\section{Conclusions}\label{sec:concl}


In this work, we have studied the dynamics of several discrete lattice models belonging to the EW and KPZ universality classes on a complete graph, with the aim of probing their infinite-dimensional limit and assessing the extent to which discrete models and their corresponding continuum descriptions yield consistent results when defined on such a highly-connected topology. We have seen that many discrete lattice models in the EW and KPZ universality classes exhibit a trivial dynamics when defined on a complete graph, in the sense that they generate asymptotically flat interfaces, consistent with the behavior displayed by the continuum equations defining these universality classes \cite{Marcos2026}. The RSOS and BD models likewise produce interfaces that are essentially flat, but their underlying dynamics remains nontrivial.


Specifically, both the RSOS and lattice BD models display several features that resemble those found for the continuum equations. First, in all cases the early-time dynamics can be characterized as a RD-like transient, during which the squared roughness grows as $w^2(t)\sim t$. Moreover, the roughness eventually reaches a time-independent regime. The time power spectrum of the spatially averaged height also exhibits a shared scaling behavior, $S(\omega)\sim \omega^{-2}$, which is consistent with linear behavior (that of, e.g., the EW equation). In addition, the two-time autocorrelation functions show essentially trivial aging: temporal correlations decay exponentially toward zero and depend only on the time difference $t-t_0$, indicating that they are effectively time-translation invariant.

Despite these similarities with the EW and KPZ equations on a complete graph, important differences arise when examining the fluctuations and the long-time dynamics. In both discrete models, the rescaled height fluctuations display a pronounced left tail, in contrast with the Gaussian fluctuations observed in the continuum equations. This feature reflects the presence of nodes that remain significantly lagging behind in the growth process while others accumulate near the maximum height. Such lagging nodes play an important role in shaping the dynamics on a fully connected substrate. 

And in turn, although the RSOS and BD models share several qualitative features, their behavior on a complete graph also differs in significant ways. In the RSOS model the saturation value of the roughness decreases with system size, qualitatively resembling the behavior observed in the continuum equations. Furthermore, the RSOS model exhibits essentially a single growth regime, namely, the RD short-time behavior of the continuum equations. In contrast, the BD model displays a qualitatively different behavior as, besides the short RD-like transient, it presents an additional ultrafast growth regime while, at long times, its time-independent roughness increases with system size. The ultrafast transient is characteristic of the BD dynamics and is already observed on regular lattices before the onset of the genuine asymptotic growth regime. Taken together, these differences show that the RSOS and BD models on a complete graph do not reproduce the scaling behavior generated by the EW and KPZ equations on the same substrate.


In summary, extending the deposition rules of these discrete lattice models to a substrate with maximal connectivity strongly affects the system dynamics, particularly the mechanisms that mimic the relaxation induced by the Laplacian term in the continuum equations. As a consequence, while the RSOS model retains some qualitative similarities with the continuum description, its saturation regime already shows noticeable deviations, whereas the BD model exhibits markedly different behavior, with dynamics that depart substantially from those exhibited by the EW and KPZ equations on a complete graph. Generally, we expect that, when formulated on a complete graph, the analysis of the many existing additional discrete growth models in the EW and/or KPZ universality classes (see e.g.\ \cite{Barabasi1995,Meakin1998,Oliveira2012,Alves2013,Oliveira2013,HalpinHealy2015,Carrasco2018}) would likewise yield a lack of scaling (in the $N\to\infty$ limit), as in the continuum equations, or perhaps trivial behavior, or else failure of universality, as assessed here for the various models covered in Sec.\ \ref{sec:models}. Naturally, further work is required to validate these expectations.


Interpreting our present results in relation to the upper critical dimension of the KPZ universality class, our simulations (together with the theoretical arguments for the additional discrete lattice models discussed in Sec.\ \ref{sec:models}) suggest that, in the effectively infinite-dimensional limit represented by a complete graph, the nonlinearity characteristic of the KPZ class becomes irrelevant, and the resulting behavior coincides with that expected for the EW class. However, this result, together with those obtained from the continuous EW and KPZ equations on the same substrate, does not allow one yet to unambiguously conclude whether the upper critical dimension of the KPZ universality class is finite or infinite. Possibly the simplest scenario could be that the KPZ fixed point flows to infinity in the infinite-dimensional limit. This would imply that the upper critical dimension of the KPZ universality class is infinite and such that no logarithmic corrections occur in that limit. However, as discussed in detail in \cite{Marcos2026} from the renormalization-group perspective, this is just one among other alternative possibilities, all of which remain to be explored.


Our results show that, on a complete graph, the RSOS and BD models yield roughness dynamics and rescaled interface-fluctuation statistics that differ both from each other and from those exhibited by the EW and KPZ equations on the same substrate. This does not imply that either microscopic model ceases to belong to the KPZ universality class. Above the upper critical dimension, universality is not guaranteed to extend to all observables, as discussed in Sec.~\ref{sec:observables} for the fluctuations of the EW equation and the RDSR model. We have explicitly verified that this occurs on a complete graph: for instance, while the fluctuations of the EW equation are Gaussian~\cite{Marcos2026}, those of the RDSR model follow a symmetric bimodal distribution with heavy tails; see ~\ref{appendix_RDSR}. Therefore, if the complete graph is regarded as an effective infinite-dimensional limit in which the KPZ nonlinearity becomes irrelevant and the system behaves as for $d>d_u$, then different microscopic models associated with the KPZ class, and the KPZ equation itself, need not display identical roughness scaling or fluctuation statistics.

\section*{Acknowledgments}

We thank the anonymous referees for their valuable comments and suggestions. This work was partially supported by Ministerio de Ciencia, Innovaci\'on y Universidades (Spain), Agencia Estatal de Investigaci\'on (AEI, Spain, 10.13039/501100011033), and European Regional Development Fund (ERDF, A way of making Europe) through Grant No.\ PID2021-123969NB-I00. The authors also acknowledge financial support through Grants No.\ PID2024-156352NB-I00 and No.\ PID2024-159024NB-C21, funded by MCIU/AEI/10.13039/501100011033/FEDER, UE and from Grant No. GR24022 funded by the Junta de Extremadura (Spain) and by European Regional Development Fund (ERDF) “A way of making Europe”. We have run our simulations in the computing facilities of the Instituto de Computaci\'{o}n Cient\'{\i}fica Avanzada de Extremadura (ICCAEx). We thank Timothy Halpin-Healy for helpful comments.

\begin{appendix}
\section{Exact roughness and rescaled height fluctuations for RDSR on a complete graph}
\label{appendix_RDSR}

We consider the RDSR model on the complete graph, starting from the flat configuration $h_i(0)=0$ for all nodes. At each deposition event, a vertex is selected uniformly at random. If one or more of its neighbours have a lower height, the incoming particle relaxes to a neighbour of minimum height; otherwise, it is deposited on the selected vertex. Since every pair of distinct vertices is connected in a complete graph, this rule implies that every particle is ultimately deposited on a vertex of current global minimum height. As a consequence, the dynamics of the RDSR model is fully deterministic, and different simulations performed for the same system size always yield identical results.

\subsection{Exact evolution of the roughness}

\begin{figure}[t]
    \centering
    \includegraphics[width=0.6\linewidth]{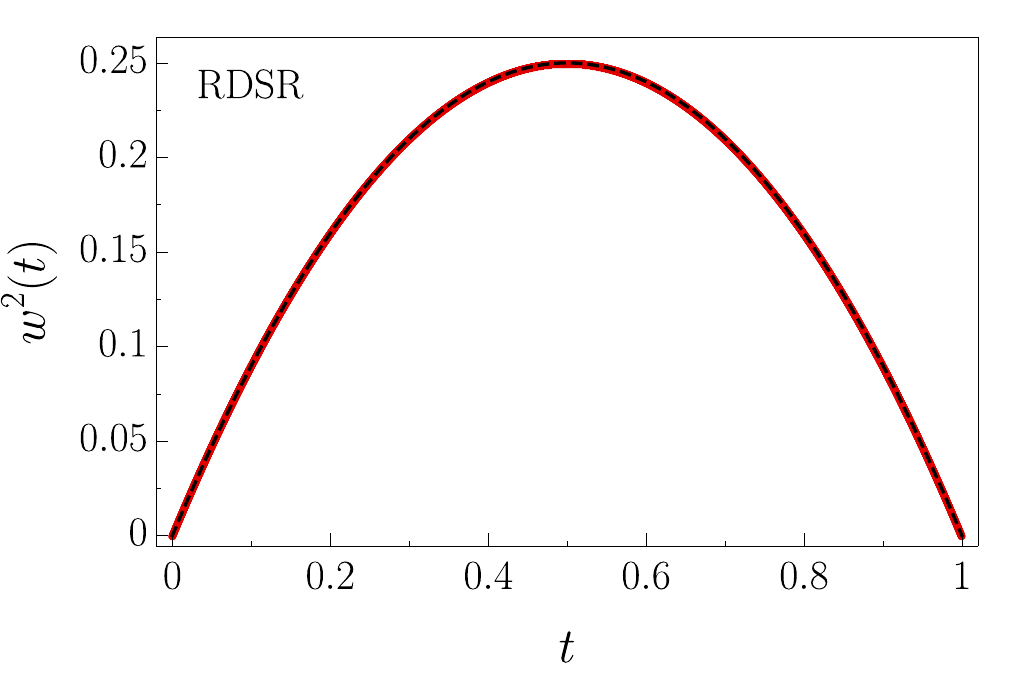}
    \caption{Time evolution of the square of the roughness, $w^2(t)$, as a function of $t$ for the RDSR model. Data is shown for system size $N=10^6$. The dashed black line corresponds to the theoretical prediction given by Eq.~\eqref{eq:rugosidad_rdsr}.}
    \label{fig:rugosidad_rdsr}
\end{figure}

Let $n$ denote the total number of deposited particles. The time is therefore given by $t=n/N$. Writing the number of deposition events as
\begin{equation}
    n=qN+r,
    \qquad
    q=\left\lfloor\frac{n}{N}\right\rfloor,
    \qquad
    0\le r<N,
\end{equation}
and defining the phase within the current monolayer as
\begin{equation}
    x=\frac{r}{N}\equiv \{t\},
\end{equation}
one finds that the surface can contain only two consecutive heights. More precisely, after $n$ deposition events,
\begin{equation}
h_i=
\left\{
\begin{array}{ll}
q, & \mbox{for } N-r=N(1-x) \mbox{ vertices}, \cr
q+1, & \mbox{for } r=Nx \mbox{ vertices}.
\end{array}
\right.
\label{eq:rdsr_alturas}
\end{equation}
This follows directly from the global-minimum deposition property: no vertex can receive
a particle at height $q+1$ before all vertices at height $q$ have been filled. The mean height is therefore
\begin{equation}
    \overline{h}(t)
    =\frac{1}{N}\sum_{i=1}^{N}h_i
    =(1-x)q+x(q+1)
    =q+x,
\end{equation}
and the squared roughness is
\begin{equation}
    w^2(t)=\frac{1}{N}\sum_{i=1}^{N}
    \left[h_i-\overline{h}(t)\right]^2
    =(1-x)x^2+x(1-x)^2=x(1-x).
    \label{eq:rugosidad_rdsr}
\end{equation}
Equation~\eqref{eq:rugosidad_rdsr} is exact and independent of the system size $N$. The roughness is periodic in monolayer time, with period one: it vanishes whenever a monolayer is completed and reaches its maximum at half filling. For even $N$, this maximum is exactly $w_{\max}=1/2$. Thus, unlike RDSR on finite-dimensional lattices, the complete-graph dynamics does not display a nontrivial kinetic-roughening regime: the width remains bounded and undergoes deterministic layer-by-layer oscillations. Figure~\ref{fig:rugosidad_rdsr} compares the simulation data with Eq.~\eqref{eq:rugosidad_rdsr}. The dashed black line represents the exact result. As can be seen in this figure, the agreement is excellent.

\subsection{Rescaled height fluctuations}

Following the definition of the rescaled height fluctuations given in Eq.~\eqref{eq:chi}, and considering a fixed phase $x$ of monolayer growth, there are only two possible height values and, consequently, only two possible values of the rescaled fluctuation:
\begin{equation}
\chi_-=
-\sqrt{\frac{x}{1-x}},
\qquad
\chi_+=
\sqrt{\frac{1-x}{x}}.
\label{eq:rdsr_dos_chi}
\end{equation}
The corresponding probabilities are $1-x$ and $x$, respectively. Therefore, at a fixed phase $x$, the fluctuation distribution is exactly
\begin{equation}
    P(\chi\mid x)
    =(1-x)\,
    \delta\left(
        \chi+\sqrt{\frac{x}{1-x}}
    \right)
    +x\,
    \delta\left(
        \chi-\sqrt{\frac{1-x}{x}}
    \right).
    \label{eq:rdsr-fixed-phase-pdf}
\end{equation}
It is important to stress that Eq.~\eqref{eq:rdsr-fixed-phase-pdf} is a two-point distribution, not a Gaussian. It has zero mean and unit variance by construction,
\begin{equation}
    \langle\chi\rangle_x=0,
    \qquad
    \langle\chi^2\rangle_x=1.
\end{equation}
At the exactly flat configurations, $x=0$, one has $w=0$, and the normalized variable in Eq.~\eqref{eq:chi} is therefore undefined.

The distribution obtained at a single fixed phase $x$ consists of only two discrete contributions and therefore does not define a continuous curve. A continuous distribution can instead be constructed by pooling the rescaled height fluctuations over all nonflat phases within a monolayer, assigning equal statistical weight to each deposition time. For a finite system of size $N$, the accessible phases are $x=r/N$, with $r=1,\ldots,N-1$. The resulting mixed distribution is obtained by averaging the two-point distribution associated with each of these phases
\begin{equation}
P_{\mathrm{mix}}^{(N)}(\chi)=
\frac{1}{N-1}
\sum_{r=1}^{N-1}
\Bigg[
\left(1-\frac{r}{N}\right)
\delta\left(
\chi+\sqrt{\frac{r}{N-r}}
\right)
+\frac{r}{N}
\delta\left(
\chi-\sqrt{\frac{N-r}{r}}
\right)
\Bigg].
\label{eq:rdsr-finite-N-mixed-pdf}
\end{equation}
Thus, the finite-size distribution remains a discrete set of peaks. In the limit $N\to\infty$, the phase becomes continuous and Eq.~\eqref{eq:rdsr-finite-N-mixed-pdf} converges, in the distributional sense, to
\begin{equation}
    P_{\mathrm{mix}}(\chi)
    =\int_0^1 P(\chi\mid x)\,dx.
    \label{eq:rdsr-continuum-mixture}
\end{equation}
For $\chi>0$, only the upper-height branch contributes. From Eq.~\eqref{eq:rdsr_dos_chi},
\begin{equation}
    \chi=\sqrt{\frac{1-x}{x}},
\end{equation}
which can be inverted as
\begin{equation}
    x(\chi)=\frac{1}{1+\chi^2}.
\end{equation}
The change of variables gives
\begin{equation}
    \left|\frac{dx}{d\chi}\right|
    =\frac{2\chi}{(1+\chi^2)^2},
    \qquad \chi>0.
\end{equation}
Multiplying this Jacobian by the statistical weight $x(\chi)$ of the upper-height vertices yields
\begin{equation}
    P_{\mathrm{mix}}(\chi)
    =x(\chi)\left|\frac{dx}{d\chi}\right|
    =\frac{2\chi}{(1+\chi^2)^3},
    \qquad \chi>0.
\end{equation}
The lower-height branch gives the symmetric result for $\chi<0$. Hence,
\begin{equation}
    P_{\mathrm{mix}}(\chi)
    =\frac{2|\chi|}{(1+\chi^2)^3}.
    \label{eq:rdsr-mixed-pdf}
\end{equation}
The distribution is normalized, symmetric, and has unit variance:
\begin{equation}
    \int_{-\infty}^{\infty}P_{\mathrm{mix}}(\chi)\,d\chi=1,
    \qquad
    \langle\chi\rangle=0,
    \qquad
    \langle\chi^2\rangle=1.
\end{equation}
Its skewness vanishes, whereas its kurtosis diverges:
\begin{equation}
S=\frac{\langle\chi^3\rangle}{\langle\chi^2\rangle^{3/2}}=0,\qquad K=\frac{\langle\chi^4\rangle}{\langle\chi^2\rangle^2}=\infty.
\end{equation}
It is manifestly non-Gaussian. In particular, its asymptotic behaviour is algebraic,
\begin{equation}
    P_{\mathrm{mix}}(\chi)
    \simeq 2|\chi|^{-5},
    \qquad |\chi|\gg 1.
    \label{eq:rdsr-tail}
\end{equation}
These tails originate from phases close to the completion or initiation of a monolayer. 



Finite-size effects are particularly transparent in this representation. Since the smallest nonzero phase is $x_{\min}=1/N$, the largest accessible fluctuation is
\begin{equation}
    |\chi|_{\max}=\sqrt{N-1}.
    \label{eq:rdsr-finite-size-cutoff}
\end{equation}
However, for the range of fluctuations typically displayed, typically $|\chi|\lesssim 5$, the formal cutoff $|\chi|_{\max}=\sqrt{N-1}$ lies outside the relevant observation window and does not constitute the dominant finite-size effect. The deviations from the continuum prediction arise primarily from the discrete set of accessible phases, $x=r/N$. Since consecutive phases are separated by $\Delta x=1/N$, the corresponding spacing between allowed fluctuation values is approximately
\begin{equation}
\Delta\chi
\simeq
\left|\frac{d\chi}{dx}\right|\frac{1}{N}
=
\frac{(1+\chi^2)^2}{2N|\chi|}.
\end{equation}
This spacing decreases as $N^{-1}$, but increases rapidly with $|\chi|$, making the discrete structure particularly visible in the tails. In the binned simulation data, this discreteness appears as families of approximately straight and parallel segments that depart from the smooth large-$N$ prediction. As the system size increases, the spacing between the allowed values decreases, these segments become shorter and less pronounced, and the empirical distribution progressively approaches the theoretical curve in Eq.~\eqref{eq:rdsr-mixed-pdf}.

\begin{figure}[t]
    \centering
    \includegraphics[width=0.6\linewidth]{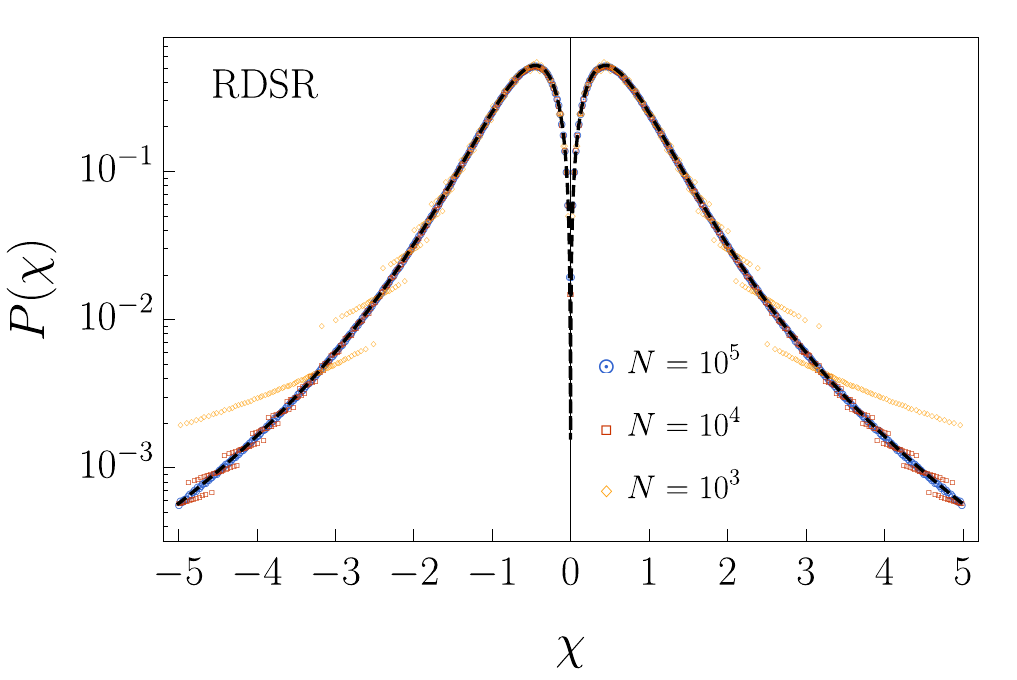}
    \caption{ Histogram of the rescaled height fluctuations $\chi$ [see Eq.~\eqref{eq:chi}] for the RDSR model. Blue circles, red squares, and orange diamonds correspond to $N=10^5$, $N=10^4$, and $N=10^3$, respectively. The dashed black line corresponds exactly to the distribution given by Eq.~\eqref{eq:rdsr-mixed-pdf}.}
    \label{fig:rdsr_chi}
\end{figure}

This convergence is illustrated in Fig.~\ref{fig:rdsr_chi}, where results for three system sizes, $N=10^3$, $N=10^4$, and $N=10^5$, are shown together with the large-$N$ theoretical prediction given in Eq.~\eqref{eq:rdsr-mixed-pdf}. The agreement improves systematically as the system size increases. For $N=10^3$, finite-size effects remain clearly visible in the tails. By contrast, for $N=10^5$, the finite-size peak structure is substantially reduced, and the simulation data follow the continuum prediction much more closely.

\section{RSOS dynamics with $m=1$}\label{appendix_RSOS_m1}

As briefly explained in Section~\ref{sec:models}, the deposition rule of the RSOS model for the $m=1$ case generates trivially flat surfaces on a fully connected graph, in the sense that the roughness is bounded and independent of the system size, and therefore does not exhibit FV scaling. Nevertheless, it is still of interest to analyze the oscillations that appear when measuring the roughness in this model.

In the present case, nodes can only take two values: that of the lower layer or that of the upper layer. Thus, the growth of the mean front position $\bar{h}(t)$ can be written as $\bar{h}(t)=n+f$, where $n$ is an integer and $f\in(0,1)$. Under these conditions the roughness is $w^2=f(1-f)$, implying $w^2\leq 1/4$, with the maximum value attained when half of the nodes are in the lower state and the other half in the upper state, i.e. $f=1/2$. The value of the roughness is completely determined by the evolution of the mean front position. In this sense, the interface exhibits a collective evolution with layer-by-layer growth, a type of dynamics also observed in both, discrete models and experiments of epitaxial surface growth~\cite{Kallabis1997,Krug1997,Michely2004}.

\begin{figure}[!t]
\centering
\includegraphics[width=0.6\textwidth]{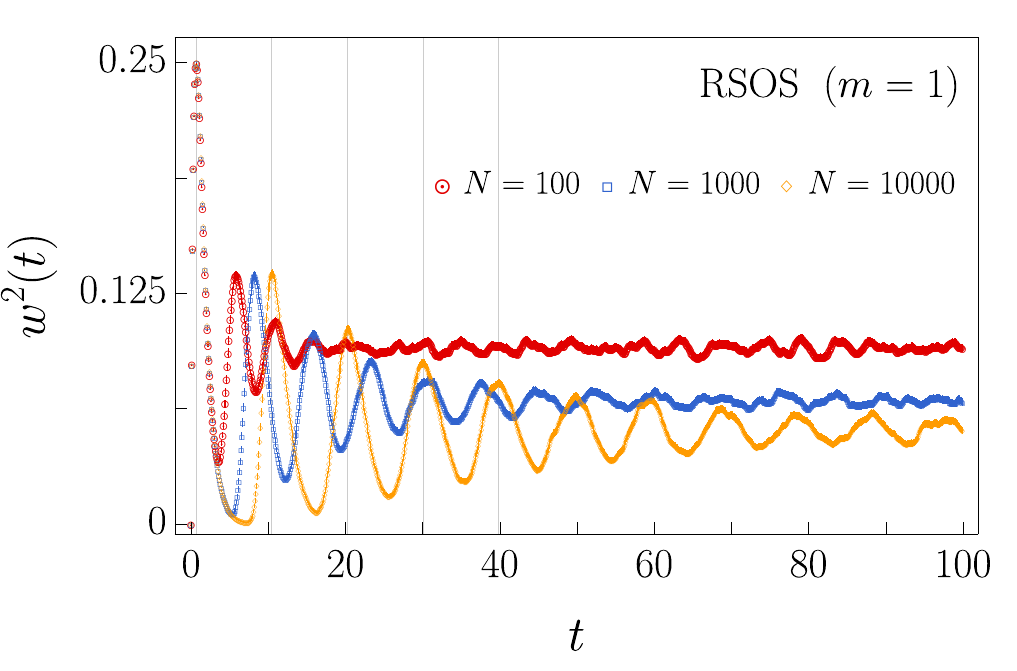}
\caption{Squared, $w^2(t)$, as a function of $t$ for the RSOS model. Data are shown for different system sizes $N$ (see legend) and $m=1$. The gridlines on the $x$-axis indicate the predicted positions of the maxima of the squared roughness for the case $N=10000$, given by $t = 0.693 + n\,T_N$, where $T_N$ is given by Eq.~\ref{eq:periodo_RSOS_m1}. The offset $0.693$ corresponds to the typical time required to reach the first maximum ($\sum_{k=0}^{N/2}\frac{1}{N-k}\approx 0.693$).}
\label{fig:w_RSOS_m1}
\end{figure}

Figure~\ref{fig:w_RSOS_m1} shows the time evolution of the squared roughness, $w^2(t)$, for the RSOS model with $m=1$ and different system sizes. The roughness can be clearly seen to exhibit oscillations, with a period that depends on the system size. Initially, the roughness increases very rapidly up to its maximum value of $0.25$, after which it decreases and enters an oscillatory regime that becomes longer-lived for larger system sizes.
The period of these oscillations can be calculated by considering the dynamics of the model. Let $k$ be the number of nodes in the upper layer. Then there are $N-k$ nodes in the lower layer. The system only advances when a node in the lower layer is selected and moves to the upper layer. The probability of selecting a node from the lower layer, and therefore of performing a deposition event, is $(N-k)/N$. Consequently, the average number of deposition attempts required for a node to move from the lower layer to the upper layer is $N/(N-k)$. Therefore, the average number of deposition attempts required to completely fill the layer is
\begin{equation}
    \sum_{k=0}^{N-1}\frac{N}{N-k}=N\left[\psi^{(0)}(N+1)+\gamma\right],
\end{equation}
where $\psi^{(0)}(x)$ is the polygamma function of order $0$ and $\gamma$ is the Euler constant ($\gamma\approx0.577$). Hence, the typical time (period) required for a system of size $N$ to complete one layer-filling cycle is
\begin{equation}
    \label{eq:periodo_RSOS_m1}
    T_N=\psi^{(0)}(N+1)+\gamma,
\end{equation}
which is precisely the period of the oscillations shown in Figure~\ref{fig:w_RSOS_m1}. Although all realizations of the noise must necessarily pass through the fully filled layer state ($f=0$), where the roughness vanishes, they do not all reach this state at the same time. As a result, the average over different noise realizations does not return strictly to zero. This effect also causes the oscillations to progressively attenuate at long times.

\begin{figure}[t]
\centering
\includegraphics[width=0.6\textwidth]{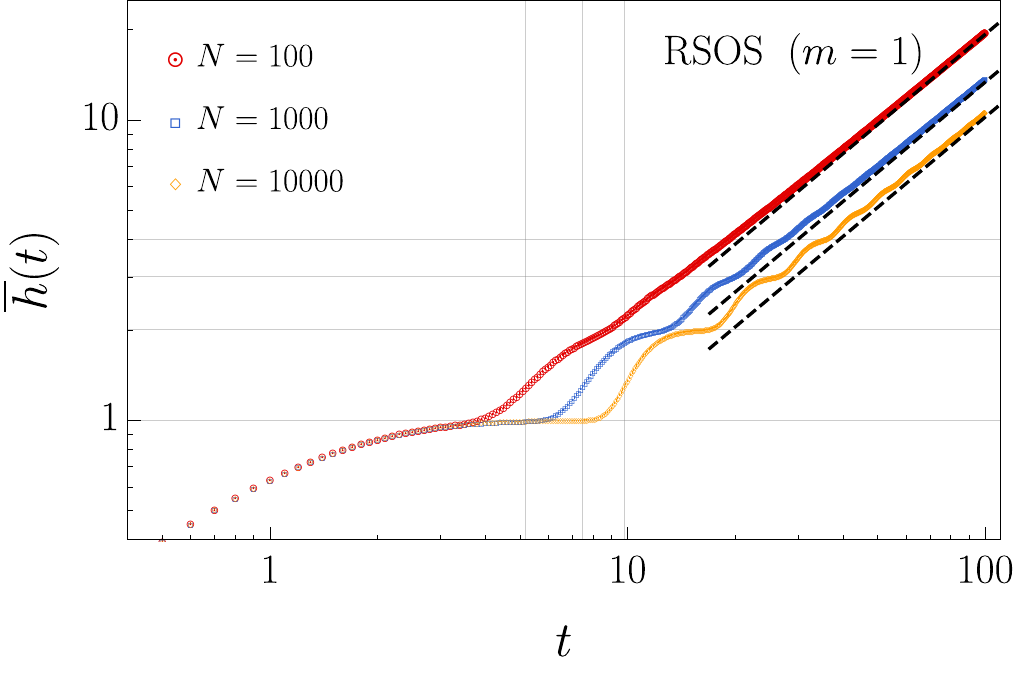}
\caption{Log--log plot of the average front position, $\bar{h}(t)$, as a function of $t$ for the RSOS model. Data are shown for different system sizes $N$ (see legend) and $m=1$. The dashed black lines correspond to the prediction given by Eq.~\eqref{eq:h_pred_RSOS_m1}. The gridlines on the $y$ axis mark the layer-filling values $1,2,3,$ and $4$, while those on the $x$ axis indicate the typical periods $T_N$ given by Eq.~\eqref{eq:periodo_RSOS_m1}.}
\label{fig:h_RSOS_m1}
\end{figure}

Since $T_n$ is the typical time required to fill one layer, the mean front position should evolve in time as
\begin{equation}
    \label{eq:h_pred_RSOS_m1}
    \bar{h}(t)=\frac{t}{T_n}.
\end{equation}
Figure~\ref{fig:h_RSOS_m1} shows the time evolution of the mean front position, $\bar{h}(t)$, for the RSOS model at different system sizes and $m=1$. In this figure, the layer-by-layer growth process can be clearly observed. The different systems initially grow very rapidly as the layer begins to fill, but then require a much longer time to complete the filling of the layer. This is consistent with the fact that the last deposition events are the least likely to be accepted, causing the evolution of the system to effectively freeze for long periods of time. Again, since each layer-filling cycle takes a different amount of time, these oscillations gradually become attenuated. Finally, at long times the system reaches the growth behavior given by Eq.~\eqref{eq:h_pred_RSOS_m1}.

\begin{figure}[!t]
\centering
\includegraphics[width=0.6\textwidth]{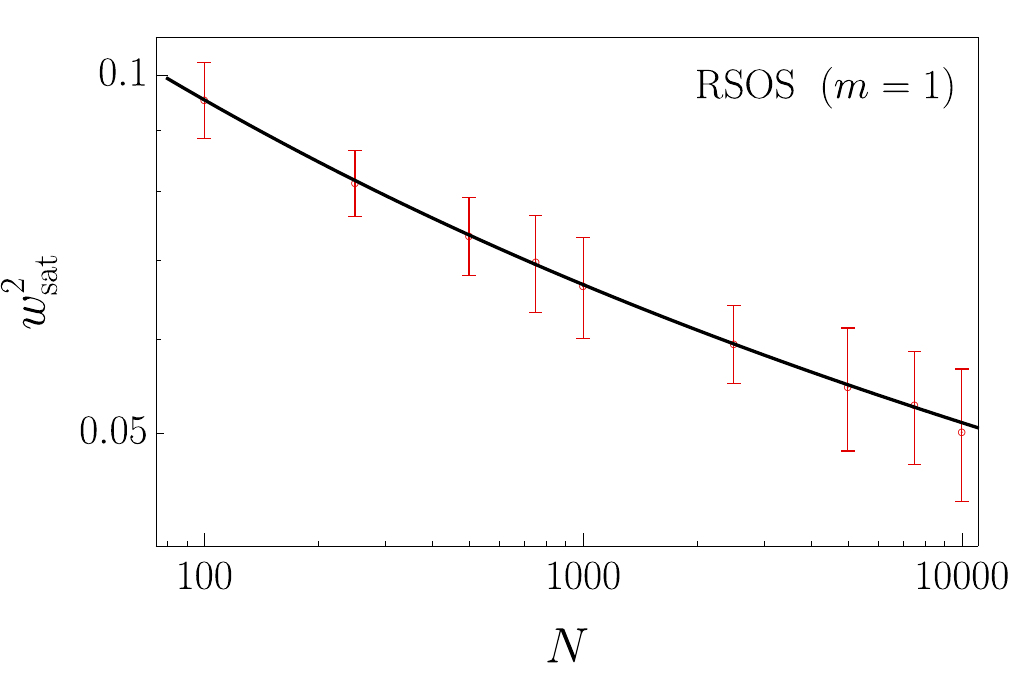}
\caption{Log--log plot of the saturation value of the squared roughness, $w^2_\mathrm{sat}$, as a function of the system size $N$ for the RSOS model. In this plot $m=1$. The solid black line corresponds to the prediction given by Eq.~\eqref{eq:w_sat_RSOS_m1}.}
\label{fig:w_sat_RSOS_m1}
\end{figure}

Finally, we analyze the saturation value of the roughness in the long-time regime. Although all noise realizations oscillate with the same average period, at long times these oscillations become completely uncorrelated, and their average over many realizations coincides with the average roughness of a single realization. This is obtained by weighting each roughness value across the different states within a cycle by the typical waiting time associated with each state. Since $1/(N-k)$ is the typical waiting time when there are $k$ nodes in the upper layer,
\begin{equation}
\label{eq:w_sat_RSOS_m1}
    w^2_{\mathrm{sat}}=\sum_{k=0}^{N-1}\frac{1}{T_N(N-k)}\frac{k}{N}\left(1-\frac{k}{N}\right)=\frac{1}{T_N}\frac{N-1}{2N}.
\end{equation}
Figure~\ref{fig:w_sat_RSOS_m1} reports the saturation value of the squared roughness, $w^2_{\mathrm{sat}}$, as a function of the system size $N$ for the RSOS model with $m = 1$, in agreement with the prediction given by equation~\eqref{eq:w_sat_RSOS_m1}.

\section{Analysis of the BD model dynamics}\label{appendix_BD_detail}

In this appendix, we provide a detailed analysis of the dynamics of the BD model on a complete graph. As shown below, the evolution is entirely governed by two key features: whenever deposition occurs at a node, its height jumps to the current maximum height of the system, whereas the maximum height itself increases only when deposition takes place at a node that is already at that maximum height.

This appendix is organized as follows. First, we analyze the dynamics of the maximum height of the system, including the typical number of nodes located at the maximum and the temporal growth of this maximum height. We then examine the time evolution of the roughness, discussing the initial RD transient, the intermediate regime of ultrafast growth, and the eventual approach to a stationary value. Next, we explain the origin of the pronounced exponential tail exhibited by the height-fluctuation distribution for large negative values of the rescaled fluctuation. Finally, we discuss the typical relaxation time of the system and provide a simple estimate of its value.

\subsection{Dynamics of the Maximum Height}
In this subsection, we analyze three closely related quantities: the average time required for the maximum height of the system to increase by one unit, the resulting temporal evolution of the maximum height, and the average number of nodes located at the maximum. Let $k(t)$ denote the number of nodes located at the current maximum height of the system. After a deposition event, two possible outcomes may occur:
\begin{itemize}
\item With probability $k/N$, deposition takes place at a node that is already at the maximum height. In this case,
$$
k \longrightarrow 1,
\qquad
h_{\mathrm{max}} \longrightarrow h_{\mathrm{max}}+1.
$$
\item With probability $1-k/N$, deposition takes place at a node below the maximum height. The selected node then reaches the current maximum, so that
$$
k \longrightarrow k+1,
\qquad
h_{\mathrm{max}} \longrightarrow h_{\mathrm{max}}.
$$
\end{itemize}
Immediately after an increase of the maximum height, the system starts from $k=1$. The number of nodes at the maximum then grows until a deposition event occurs at one of these nodes, causing the maximum height to increase and resetting $k$ to unity. We now estimate the typical time required to complete such a cycle, or equivalently, the typical time between two successive increases of the maximum height.

\begin{figure}[t]
\centering
\includegraphics[width=0.6\textwidth]{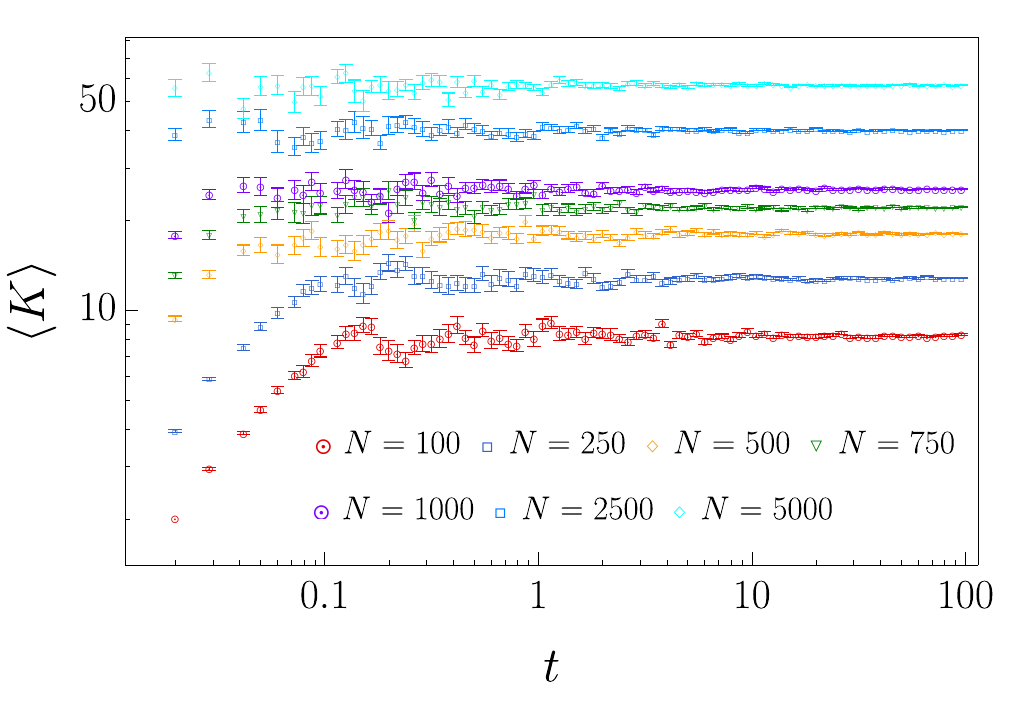}
\caption{Log--log plot of the number of sites at the maximum value, $\langle K\rangle$, as a function of $t$. Data are shown for different system sizes $N$ (see legend).}
\label{fig:k}
\end{figure}

\begin{figure}[t]
\centering
\includegraphics[width=0.6\textwidth]{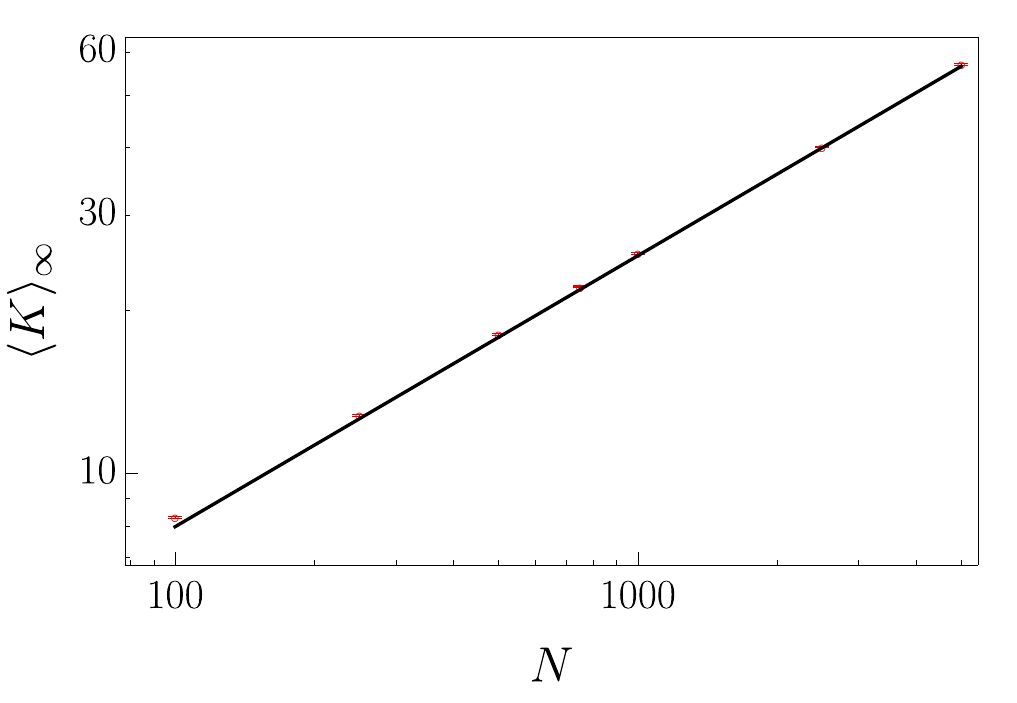}
\caption{Log--log plot of the saturation value of the number of sites at the maximum value, $\langle K\rangle_\infty$, as a function of the system size $N$. The solid black line represents the theoretical prediction given by Eq.~\eqref{eq:k_average_2}.}
\label{fig:k_inf}
\end{figure}

Starting from $k=1$, the probability that the process reaches a state with $k$ nodes at the maximum height before the maximum itself increases is
\begin{equation}
\label{eq:prob_estado_k}
\prod_{j=1}^{k-1}\frac{N-j}{N}.
\end{equation}
The typical time required for the system to return to $k=1$, or equivalently for the maximum height to increase by one unit, is given by the mean duration of a cycle. Measured in deposition events, this time is
\begin{equation}
\label{eq:tiempo_BD}
\tau_{\mathrm{cycle,BD}}=
\sum_{k=1}^{N}\frac{k}{N} k
\prod_{j=1}^{k-1}\left(\frac{N-j}{N}\right),
\end{equation}
where $k/N$ is the probability that the maximum height increases at that step and $k$ accounts for the number of nodes accumulated at the maximum, which is directly related to the number of deposition events elapsed since the beginning of the cycle. To evaluate this sum, let us first examine the product term. Taking logarithms
\begin{eqnarray}
\ln (a_k)&=& \ln \left(\prod_{j=1}^{k-1}\frac{N-j}{N}\right)= \sum_{j=1}^{k-1}\ln\left(1-\frac{j}{N}\right)\approx-\sum_{j=1}^{k-1}j/N= \nonumber\\[0.4em]
&=& -\frac{1}{N}\sum_{j=1}^{k-1}j=-\frac{1}{N}\frac{(k-1)k}{2}\approx-\frac{k^2}{2N}.
\end{eqnarray}
Therefore,
\begin{equation}
\label{eq:tiempo_BD_2}
\tau_{\mathrm{cycle,BD}}\approx\frac{1}{N}
\sum_{k=1}^{N}k^2e^{-k^2/2N}.
\end{equation}
Taking the limit $N\to\infty$, while approximating the sum by an integral,
\begin{equation}
\label{eq:tiempo_BD_3}
\tau_{\mathrm{cycle,BD}}\approx\frac{1}{N}\int_0^\infty k^2 e^{-k^2/2N}dk=\sqrt{N}\int_0^\infty u^2 e^{-u^2/2}du=\sqrt{\frac{\pi N}{2}}.
\end{equation}
One of the main consequences of this result is that the maximum height grows approximately as
\begin{equation}
\label{eq:hmax_growth}
h_\mathrm{max}(t)=\frac{tN}{\tau_{\mathrm{cycle,BD}}}=t\sqrt{\frac{2 N}{\pi}},
\end{equation}
where time $t$ is measured in deposition events per node. Finally, let us determine the average number of nodes at the maximum height, 
\begin{equation}
\label{eq:k_average}
\langle K\rangle=\frac{\sum_{k=1}^N k a_k}{\sum_{k=1}^N a_k}=\frac{\sum_{k=1}^N k e^{-k^2/2N}}{\sum_{k=1}^N e^{-k^2/2N}}.
\end{equation}
Taking the $N\to\infty$ limit once again and approximating the sums by integrals,
\begin{eqnarray}
\sum_{k=1}^N e^{-k^2/2N}&\approx& \int_0^\infty e^{-k^2/2N} dk=\sqrt{N} \int_0^\infty e^{-u^2/2}du= \sqrt{\frac{\pi N}{2}},\nonumber\\[0.4em]
\sum_{k=1}^N k e^{-k^2/2N}&\approx& \int_0^\infty k e^{-k^2/2N} dk=N \int_0^\infty u e^{-u^2/2}du=N.
\end{eqnarray}
Therefore,
\begin{equation}
\label{eq:k_average_2}
\langle K\rangle=\sqrt{\frac{2N}{\pi}}.
\end{equation}
We have verified numerically that this asymptotic behavior is reached rapidly, with only a short transient for small system sizes. Figure~\ref{fig:k} shows the temporal evolution of $\langle K\rangle$ for several values of $N$, while Fig.~\ref{fig:k_inf} displays its stationary value, $\langle K\rangle_\infty$, as a function of the system size. The solid line in Fig.~\ref{fig:k_inf} represents the theoretical prediction of Eq.~\eqref{eq:k_average_2}.

\subsection{Time Evolution of the Roughness}

In this subsection, we analyze the temporal evolution of the roughness. We first discuss the initial random-deposition transient, then the intermediate ultrafast-growth regime, and finally the approach to the stationary roughness.

For very short times and very small systems, it is possible to observe a scaling behavior of the form $w^2(t) \sim t$, as indicated by the dashed lines shown in Fig.~\ref{fig:w_BD}. Simply put, when $t < 1/\sqrt{N}$, the behavior of the system is essentially the same as that of the RD model.

During these initial stages, particles are deposited but the maximum height does not increase, since most sites still have $h_i = 0$. The probability that the maximum increases grows with time as more particles are deposited, and is given by $\frac{tN}{N} = t$. Therefore, over a time interval $dt$ the number of deposition events is $N\,dt$, and the maximum height increases on average by
\begin{equation}
\int_0^t t' N \, dt' = \frac{N t^2}{2}.
\end{equation}
Imposing the condition $\frac{N t^2}{2} \approx 1$, which corresponds to assuming that the maximum has increased and that this random deposition regime has ended, we obtain the crossover time
\begin{equation}
t \sim \frac{1}{\sqrt{N}}.
\end{equation}
This transient can only be observed in Fig.~\ref{fig:w_BD} and Fig.~\ref{fig:w_BD_colapso_2} for small system sizes, such as $N = 100$.

To analyze the ultrafast-growth regime, we distinguish between visited nodes, which have experienced at least one deposition event, and unvisited nodes, whose heights remain equal to zero. Since a given node is selected with probability $1/N$ at each deposition event, the number of events received by that node after $n=tN$ deposition attempts converges, in the large-$N$ limit, to a Poisson distribution with mean $t$,
\begin{equation}
P\bigl(K_i(t)=k\bigr)
=
e^{-t}\frac{t^k}{k!}.
\end{equation}
In particular, the probability that the node remains unvisited is
\begin{equation}
P\bigl(K_i(t)=0\bigr)=e^{-t},
\end{equation}
whereas the probability that it has been visited at least once is $1-e^{-t}$.

If, by time $t$, a given node has received exactly $k$ deposition events, their occurrence times lie in the interval $(0,t)$. Conditioned on this number of events, these times are independently and uniformly distributed over that interval,
\begin{equation}
U_1,\ldots,U_k \sim \mathrm{Uniform}(0,t).
\end{equation}
The height of the node is determined by its most recent deposition event. Therefore, the relevant time is
\begin{equation}
S_k=\max(U_1,\ldots,U_k).
\end{equation}
We now determine the probability distribution of this last-deposition time. For $0\leq s\leq t$, since the variables $U_j$ are independent and uniformly distributed over $(0,t)$,
\begin{equation}
P(S_k\leq s)
=
P(U_1\leq s,\ldots,U_k\leq s)
=
\prod_{j=1}^{k}P(U_j\leq s)
=
\left(\frac{s}{t}\right)^k.
\end{equation}
The corresponding probability density function is therefore
\begin{equation}
f_{S_k}(s)
=
\frac{d}{ds}\left(\frac{s}{t}\right)^k
=
\frac{k s^{k-1}}{t^k},
\qquad 0\leq s\leq t.
\end{equation}
The moments of this distribution are given by
\begin{equation}
\left\langle S_k^m\right\rangle
=
\int_0^t s^m f_{S_k}(s)ds
=
\frac{k}{t^k}\int_0^t s^{k+m-1}ds
=
\frac{k}{k+m}t^m.
\end{equation}
In particular, the first and second moments are
\begin{equation}
\left\langle S_k\right\rangle
=
\frac{k}{k+1}t,
\qquad
\left\langle S_k^2\right\rangle
=
\frac{k}{k+2}t^2.
\end{equation}
Let us now determine how these results translate into the node heights. In the previous subsection, we found that the maximum height grows according to Eq.~\eqref{eq:hmax_growth}. Since the most recent deposition event at node $i$ occurred at time $S_k$, its height at time $t$ is approximately equal to the maximum height at that earlier time,
\begin{equation}
h_i(t)=h_{\mathrm{max}}(S_k)=\sqrt{\frac{2N}{\pi}}S_k\equiv vS_k,
\qquad
v=\sqrt{\frac{2N}{\pi}}.
\end{equation}
Therefore, conditioned on node $i$ having received exactly $k$ deposition events by time $t$, the first and second moments of its height are
\begin{equation}
\left\langle h_i(t)\mid K_i(t)=k\right\rangle=v\left\langle S_k\right\rangle=vt\frac{k}{k+1},
\end{equation}
and
\begin{equation}
\left\langle h_i^2(t)\mid K_i(t)=k\right\rangle=v^2\left\langle S_k^2\right\rangle=v^2t^2\frac{k}{k+2}.
\end{equation}
We can now compute the mean height, $\overline{h}(t)$, and the second moment of the height distribution, $\overline{h^2}(t)$.
\begin{equation}
\overline{h}(t)=\sum_{k=1}^{\infty}P\bigl(K_i(t)=k\bigr)\left\langle h_i(t)\mid k_i(t)=k\right\rangle=vte^{-t}\sum_{k=1}^{\infty}\frac{t^k}{k!}\frac{k}{k+1},
\end{equation}
where $P\bigl(K_i(t)=k\bigr)$ is the Poisson distribution with mean $t$, as discussed above. The term $k=0$ does not contribute to the sum, since a node that has not yet been visited remains at height $h_i=0$. Evaluating the sum, we obtain
\begin{equation}
\overline{h}(t)=v t e^{-t} \left[e^t \left(1-\frac{1}{t}\right)+\frac{1}{t}\right]=v\left(t-1+e^{-t}\right).
\end{equation}
The second moment is
\begin{eqnarray}
    \overline{h^2}(t)&=&\sum_{k=1}^{\infty}P\bigl(K_i(t)=k\bigr)\left\langle h_i(t)\mid K_i(t)=k\right\rangle=v^2t^2e^{-t}\sum_{k=1}^{\infty}\frac{t^k}{k!}\frac{k}{k+2} \nonumber\\[0.4em]
    &=& v^2t^2e^{-t}\left[e^t \left(1-\frac{2}{t}+\frac{2}{t^2}\right)-\frac{2}{t^2}\right]=v^2\left(t^2-2t+2-2 e^{-t}\right),
\end{eqnarray}
and therefore the squared roughness is
\begin{equation}
\label{eq:rugosidad_BD_ultrafast}
w^2(t)=\overline{h^2}(t)-\left(\overline{h}(t)\right)^2=\frac{2N}{\pi}\left(1-e^{-2t}-2te^{-t}\right).
\end{equation}
Interestingly, the same scaling function, up to an overall prefactor and a rescaling of time, appears in the growth of correlations in a system of Brownian particles subject to simultaneous Poissonian resetting~\cite{DeMauro2026}. This similarity may point to a common underlying renewal-like mechanism, despite the different microscopic dynamics of the two models.

The effective exponent depends on time and can be defined as
\begin{equation}
\beta_{\mathrm{eff}}(t)=\frac{d\ln w^2(t)}{d\ln t}=
\frac{2t e^{-t}\left(e^{-t}+t-1\right)}{1-e^{-2t}-2te^{-t}}.
\end{equation}
In particular, at $t=1$ one finds $\beta_{\mathrm{eff}}\simeq 2.1$, which explains why an approximate behavior of the form $w^2(t)\sim t^2$ is observed in this intermediate-time regime.

Equation~\eqref{eq:rugosidad_BD_ultrafast} already predicts the long-time value of the roughness. In particular, it yields the following scaling:
\begin{equation}
\label{eq:rugosidad_BD_saturacion}
w^2_{\mathrm{st}}=\lim_{t\to\infty} w^2(t)=\frac{2N}{\pi}\approx 0.636N,
\end{equation}
in good agreement with the results shown in Fig.~\ref{fig:w_BD_colapso_2}.

The physical mechanism leading to the stationary roughness can be understood as follows. In the stationary regime, a node that has not been visited recently has a height
\begin{equation}
h_i=h_{\mathrm{max}}-\Delta_i,
\end{equation}
where $\Delta_i$ is the number of times the maximum height has increased since the most recent deposition event at that node. Therefore, the mean height can be written as
\begin{equation}
\overline{h}(t)=h_{\max}(t)-\overline{\Delta}(t),
\end{equation}
and 
\begin{equation}
h_i(t)-\overline{h}(t)=-\left[\Delta_i(t)-\overline{\Delta}(t)\right].
\end{equation}
It follows exactly that
\begin{equation}
w^2(t)=\overline{\left[h_i(t)-\overline{h}(t)\right]^2}
=\overline{\left[\Delta_i(t)-\overline{\Delta}(t)\right]^2}
=\mathrm{Var}\left[\Delta(t)\right].
\end{equation}
Thus, the roughness is determined not by the mean distance from the maximum, but by the variance of this distance. Since the typical number of deposition events between two successive visits to a given node is of order $N$, the typical height difference between that node and the current maximum scales as
\begin{equation}
\Delta h_{\mathrm{max}}
\sim
\frac{N}{\tau_{\mathrm{cycle,BD}}}
=
\sqrt{\frac{2N}{\pi}}.
\end{equation}
Consequently, the stationary roughness scales as
\begin{equation}
\label{eq:rugosidad_BD_saturacion_2}
w_{\mathrm{sat}}
\sim
\sqrt{\frac{2N}{\pi}},
\end{equation}
or, equivalently, $w_{\mathrm{sat}}^2\sim 2N/\pi$, in agreement with Eq.~\eqref{eq:rugosidad_BD_saturacion}.


\subsection{Exponential Tail of the Height-Fluctuation Distribution}

In this subsection, we explain why the rescaled height fluctuations $\chi$ develop an exponential tail for negative values in the stationary regime. A simple derivation of the stationary distribution of the rescaled height fluctuations can be obtained directly from the time of the most recent deposition event. Let $S_i(t)$ denote the time at which node $i$ was last visited before the observation time $t$. As discussed above, the maximum height grows linearly,
\begin{equation}
h_{\mathrm{max}}(t)= vt,\qquad v=\sqrt{\frac{2N}{\pi}},
\end{equation}
and the height of node $i$ is approximately
\begin{equation}
h_i(t)=h_{\mathrm{max}}\bigl(S_i(t)\bigr)= vS_i(t).
\end{equation}

It is convenient to introduce the time elapsed since the most recent visit to this node, denoted by $A_i$,
\begin{equation}
A_i(t)=t-S_i(t).
\end{equation}
As discussed above, the deposition events at a given node form a Poisson process with unit rate when time is measured in deposition events per node. Therefore, in the stationary regime, the probability that the node has not been visited during the preceding interval of duration $a$ is
\begin{equation}
\Pr\bigl(A_i>a\bigr)=e^{-a}.
\end{equation}
Hence, $A_i$ is exponentially distributed with unit rate,
\begin{equation}
P_A(a)=e^{-a},
\qquad
a\geq 0.
\end{equation}

Using $S_i(t)=t-A_i(t)$, the node height can be written as
\begin{equation}
h_i(t)=v\bigl[t-A_i(t)\bigr].
\end{equation}
Since an exponential random variable of unit rate satisfies
\begin{equation}
\langle A_i\rangle=1,\qquad\mathrm{Var}(A_i)=1,
\end{equation}
the mean height and the stationary roughness are
\begin{equation}
\overline{h}(t)=v(t-1),
\end{equation}
and
\begin{equation}
w_{\mathrm{sat}}=v.
\end{equation}
The rescaled height fluctuations are therefore
\begin{equation}
\chi=\frac{h_i-\overline{h}}{w_{\mathrm{sat}}}=1-A_i.
\end{equation}
Performing the change of variables $A_i=1-\chi$, we obtain
\begin{equation}
P(\chi)=P_A(1-\chi)\left|\frac{dA_i}{d\chi}\right|=
e^{-(1-\chi)}=e^{\chi-1},\qquad\chi\leq 1.
\end{equation}
Thus, the exponential tail of the stationary height-fluctuation distribution follows directly from the exponential distribution of the time elapsed since the most recent deposition event at a node. This is precisely the distribution shown in Fig.~\ref{fig:chi_BD}.

\subsection{Relaxation Time}

We finally address why the relaxation time is approximately independent of the system size and provide a simple estimate of its value. Since time is measured in deposition events per node, a time interval $t$ corresponds to $tN$ deposition attempts. As discussed above, the number of deposition events received by a given node up to time $t$ is, in the large-$N$ limit, Poisson distributed with mean $t$:
\begin{equation}
K_i(t)\sim\mathrm{Pois}(t).
\end{equation}
Thus, the local deposition statistics at a fixed time $t$ do not depend on $N$.

A simple estimate of the relaxation time can be obtained by assuming that the system has effectively lost memory of its initial condition once a prescribed fraction $p$ of the nodes has received at least $m$ deposition events. Accordingly, we define $p$ as the expected fraction of nodes satisfying $K_i(t)\geq m$:
\begin{equation}
p
=
P\bigl(K_i(t)\geq m\bigr).
\end{equation}
Since $K_i(t)$ follows a Poisson distribution with mean $t$, this probability is
\begin{equation}
P\bigl(K_i(t)\geq m\bigr)
=
1-e^{-t}
\sum_{k=0}^{m-1}
\frac{t^k}{k!}.
\end{equation}
The relaxation time can therefore be estimated by solving
\begin{equation}
1-e^{-t_{\mathrm{sat}}}
\sum_{k=0}^{m-1}
\frac{t_{\mathrm{sat}}^k}{k!}
=
p.
\end{equation}
Because this equation contains no explicit dependence on $N$, the resulting estimate of $t_{\mathrm{sat}}$ is the same for all system sizes.

For example, taking $p=0.99$, so that $99\%$ of the nodes are expected to have received at least $m$ deposition events, gives
\begin{equation}
t_{\mathrm{sat}}\simeq 6.64\quad (m=2),
\end{equation}
and
\begin{equation}
t_{\mathrm{sat}}\simeq 8.41\quad (m=3).
\end{equation}
These estimates are consistent with Fig.~\ref{fig:w_BD}, where the stationary regime is reached at times of order $t\simeq 7$. This suggests that the system reaches its stationary regime once almost all nodes have undergone only a few deposition events.



\end{appendix}

\newpage
\section*{References}

\bibliographystyle{iopart-num}

\begin{thebibliography}{10}
\expandafter\ifx\csname url\endcsname\relax
  \def\url#1{{\tt #1}}\fi
\expandafter\ifx\csname urlprefix\endcsname\relax\def\urlprefix{URL }\fi
\providecommand{\eprint}[2][]{\url{#2}}

\bibitem{Kardar1986}
Kardar M, Parisi G and Zhang Y~C 1986 {\em Phys. Rev. Lett.\/} {\bf 56} 889 \urlprefix\url{https://link.aps.org/doi/10.1103/PhysRevLett.56.889}

\bibitem{Barabasi1995}
Barab{\'{a}}si A~L and Stanley H~E 1995 {\em Fractal Concepts in Surface Growth\/} (Cambridge: Cambridge University Press)

\bibitem{Krug1997}
Krug J 1997 {\em Adv. Phys.\/} {\bf 46} 139 \urlprefix\url{https://doi.org/10.1080/00018739700101498}

\bibitem{Takeuchi2018}
Takeuchi K~A 2018 {\em Physica A\/} {\bf 504} 77 \urlprefix\url{https://doi.org/10.1016/j.physa.2018.03.009}

\bibitem{Caballero2020}
Caballero F and Cates M~E 2020 {\em Phys. Rev. Lett.\/} {\bf 124} 240604 \urlprefix\url{https://doi.org/10.1103/PhysRevLett.124.240604}

\bibitem{Sieberer2025}
Sieberer L~M, Buchhold M, Marino J and Diehl S 2025 {\em Rev. Mod. Phys.\/} {\bf 97} 025004 \urlprefix\url{http://dx.doi.org/10.1103/revmodphys.97.025004}

\bibitem{Edwards1982}
Edwards S~F and Wilkinson D~R 1982 {\em Proc. Roy. Soc. London. A\/} {\bf 381} 17--31 \urlprefix\url{https://doi.org/10.1098/rspa.1982.0056}

\bibitem{Kardar2007}
Kardar M 2007 {\em Statistical Physics of Fields\/} (Cambridge: Cambridge University Press)

\bibitem{Alves2014}
Alves S~G, Oliveira T~J and Ferreira S~C 2014 {\em Phys. Rev. E\/} {\bf 90} 020103(R) \urlprefix\url{https://link.aps.org/doi/10.1103/PhysRevE.90.020103}

\bibitem{Oliveira2022}
Oliveira T~J 2022 {\em Phys. Rev. E\/} {\bf 106} L062103 \urlprefix\url{https://link.aps.org/doi/10.1103/PhysRevE.106.L062103}

\bibitem{HalpinHealy2025}
Halpin-Healy T 2025 {\em Phys. Rev. E\/} {\bf 111} 014147 \urlprefix\url{http://dx.doi.org/10.1103/PhysRevE.111.014147}

\bibitem{Halpin-Healy1990}
Halpin-Healy T 1990 {\em Phys. Rev. A\/} {\bf 42} 711--722 \urlprefix\url{https://link.aps.org/doi/10.1103/PhysRevA.42.711}

\bibitem{Bouchaud1993}
Bouchaud J~P and Cates M~E 1993 {\em Phys. Rev. E\/} {\bf 47} R1455--R1458 \urlprefix\url{https://link.aps.org/doi/10.1103/PhysRevE.47.R1455}

\bibitem{Doherty1994}
Doherty J~P, Moore M~A, Kim J~M and Bray A~J 1994 {\em Phys. Rev. Lett.\/} {\bf 72} 2041--2044 \urlprefix\url{https://link.aps.org/doi/10.1103/PhysRevLett.72.2041}

\bibitem{Colaiori2001}
Colaiori F and Moore M~A 2001 {\em Phys. Rev. Lett.\/} {\bf 86} 3946--3949 \urlprefix\url{https://link.aps.org/doi/10.1103/PhysRevLett.86.3946}

\bibitem{Kloss2012}
Kloss T, Canet L and Wschebor N 2012 {\em Phys. Rev. E\/} {\bf 86} 051124 \urlprefix\url{https://doi.org/10.1103/PhysRevE.86.051124}

\bibitem{Kloss2014}
Kloss T, Canet L, Delamotte B and Wschebor N 2014 {\em Phys. Rev. E\/} {\bf 89} 022108 \urlprefix\url{https://link.aps.org/doi/10.1103/PhysRevE.89.022108}

\bibitem{Kloss2014-2}
Kloss T, Canet L and Wschebor N 2014 {\em Phys. Rev. E\/} {\bf 90} 062133 \urlprefix\url{https://link.aps.org/doi/10.1103/PhysRevE.90.062133}

\bibitem{Canet2025}
Canet L 2025 {\em J. Stat. Mech.\/} {\bf 2025} 124003 \urlprefix\url{https://dx.doi.org/10.1088/1742-5468/ae1e75}

\bibitem{Castellano1998}
Castellano C, Marsili M and Pietronero L 1998 {\em Phys. Rev. Lett.\/} {\bf 80} 3527--3530 \urlprefix\url{https://link.aps.org/doi/10.1103/PhysRevLett.80.3527}

\bibitem{Castellano1998-2}
Castellano C, Gabrielli A, Marsili M, Mu\~noz M~A and Pietronero L 1998 {\em Phys. Rev. E\/} {\bf 58} R5209--R5212 \urlprefix\url{https://link.aps.org/doi/10.1103/PhysRevE.58.R5209}

\bibitem{Tauber2014}
T\"auber U~C 2014 {\em Critical Dynamics\/} (Cambridge: Cambridge University Press)

\bibitem{Marcos2026}
Marcos J~M, Meléndez J~J, Cuerno R and Ruiz-Lorenzo J~J 2026 {\em SciPost Physics\/} {\bf 21} ISSN 2542-4653 \urlprefix\url{http://dx.doi.org/10.21468/SciPostPhys.21.4.087}

\bibitem{Marcos2025}
Marcos J~M, Meléndez J~J, Cuerno R and Ruiz-Lorenzo J~J 2025 {\em J. Stat. Mech.: Theor. Exp.\/} {\bf 2025} 083203 \urlprefix\url{http://dx.doi.org/10.1088/1742-5468/adf295}

\bibitem{Saberi_2013}
Saberi A~A 2013 {\em EPL\/} {\bf 103} 10005 \urlprefix\url{https://dx.doi.org/10.1209/0295-5075/103/10005}

\bibitem{Oliveira2021}
Oliveira T~J 2021 {\em EPL\/} {\bf 133} 28001 \urlprefix\url{https://dx.doi.org/10.1209/0295-5075/133/28001}

\bibitem{Derrida1998}
Derrida B 1998 {\em Phys. Rep.\/} {\bf 301} 65--83 \urlprefix\url{https://doi.org/10.1016/S0370-1573(98)00006-4}

\bibitem{Spohn1991}
Spohn H 1991 {\em Large Scale Dynamics of Interacting Particles\/} (Berlin: Springer)

\bibitem{Kim1989}
Kim J~M and Kosterlitz J~M 1989 {\em Phys. Rev. Lett.\/} {\bf 62} 2289–2292 \urlprefix\url{http://dx.doi.org/10.1103/PhysRevLett.62.2289}

\bibitem{Kim2014}
Kim S~W and Kim J~M 2014 {\em J. Stat. Mech.: Theor. Exp.\/} {\bf 2014} P07005 \urlprefix\url{https://dx.doi.org/10.1088/1742-5468/2014/07/P07005}

\bibitem{Prhofer2000}
Pr\"{a}hofer M and Spohn H 2000 {\em Phys. Rev. Lett.\/} {\bf 84} 4882–4885 \urlprefix\url{http://dx.doi.org/10.1103/PhysRevLett.84.4882}

\bibitem{Johansson2000}
Johansson K 2000 {\em Comm. Math. Phys.\/} {\bf 209} 437–476 \urlprefix\url{http://dx.doi.org/10.1007/s002200050027}

\bibitem{Mello2001}
Mello B~A, Chaves A~S and Oliveira F~A 2001 {\em Phys. Rev. E\/} {\bf 63} \urlprefix\url{http://dx.doi.org/10.1103/PhysRevE.63.041113}

\bibitem{Alves2016-2}
Alves W~S, Rodrigues E~A, Fernandes H~A, Mello B~A, Oliveira F~A and Costa I~V~L 2016 {\em Phys. Rev. E\/} {\bf 94} \urlprefix\url{http://dx.doi.org/10.1103/PhysRevE.94.042119}

\bibitem{Rodrigues2015}
Rodrigues E, Oliveira F and Mello B 2015 {\em Acta Phys. Pol. B\/} {\bf 46} 1231 \urlprefix\url{http://dx.doi.org/10.5506/APhysPolB.46.1231}

\bibitem{AaroReis2004}
Aarão~Reis F~D~A 2004 {\em Phys. Rev. E\/} {\bf 69} \urlprefix\url{http://dx.doi.org/10.1103/PhysRevE.69.021610}

\bibitem{Baxter2016}
Baxter R~J 1989 {\em Exactly solved models in statistical mechanics\/} (London: Academic Press)

\bibitem{Newman1999}
Newman M~E~J and Barkema G~T 1999 {\em Monte Carlo Methods in Statistical Physics\/} (Oxford: Oxford University Press)

\bibitem{Peterson2011}
Peterson J 2011 {\em Stochastic Processes and their Applications\/} {\bf 121} 609–629 ISSN 0304-4149 \urlprefix\url{http://dx.doi.org/10.1016/j.spa.2010.11.003}

\bibitem{Xue2017}
Xue X and Pan Y 2017 {\em Journal of Statistical Physics\/} {\bf 169} 951–971 ISSN 1572-9613 \urlprefix\url{http://dx.doi.org/10.1007/s10955-017-1898-4}

\bibitem{OttinoLffler2017}
Ottino-L\"{o}ffler B, Scott J~G and Strogatz S~H 2017 {\em Phys. Rev. E\/} {\bf 96} ISSN 2470-0053 \urlprefix\url{http://dx.doi.org/10.1103/PhysRevE.96.012313}

\bibitem{Guo2013}
Guo D, Trajanovski S, van~de Bovenkamp R, Wang H and Van~Mieghem P 2013 {\em Phys. Rev. E\/} {\bf 88} ISSN 1550-2376 \urlprefix\url{http://dx.doi.org/10.1103/PhysRevE.88.042802}

\bibitem{Lipowski2022}
Lipowski A and Lipowska D 2022 {\em Phys. Rev. E\/} {\bf 105} ISSN 2470-0053 \urlprefix\url{http://dx.doi.org/10.1103/PhysRevE.105.024119}

\bibitem{Azhari2022}
Azhari and Muslim R 2022 {\em International Journal of Modern Physics C\/} {\bf 34} ISSN 1793-6586 \urlprefix\url{http://dx.doi.org/10.1142/S0129183123500882}

\bibitem{Fronczak2017}
Fronczak A and Fronczak P 2017 {\em Phys. Rev. E\/} {\bf 96} ISSN 2470-0053 \urlprefix\url{http://dx.doi.org/10.1103/PhysRevE.96.012304}

\bibitem{Sood2008}
Sood V, Antal T and Redner S 2008 {\em Phys. Rev. E\/} {\bf 77} ISSN 1550-2376 \urlprefix\url{http://dx.doi.org/10.1103/PhysRevE.77.041121}

\bibitem{Huang2018}
Huang W, Hou P, Wang J, Ziff R~M and Deng Y 2018 {\em Phys. Rev. E\/} {\bf 97} ISSN 2470-0053 \urlprefix\url{http://dx.doi.org/10.1103/PhysRevE.97.022107}

\bibitem{Takeuchi2012}
Takeuchi K~A and Sano M 2012 {\em J. Stat. Phys.\/} {\bf 147} 853 \urlprefix\url{https://dx.doi.org/10.1007/s10955-012-0503-0}

\bibitem{Henkel2012}
Henkel M, Noh J~D and Pleimling M 2012 {\em Phys. Rev. E\/} {\bf 85} 030102(R) \urlprefix\url{https://doi.org/10.1103/PhysRevE.85.030102}

\bibitem{DeNardis2017}
De~Nardis J, Le~Doussal P and Takeuchi K~A 2017 {\em Physical Review Letters\/} {\bf 118} ISSN 1079-7114 \urlprefix\url{http://dx.doi.org/10.1103/PhysRevLett.118.125701}

\bibitem{Rthlein2006}
R\"{o}thlein A, Baumann F and Pleimling M 2006 {\em Phys. Rev. E\/} {\bf 74} 061604 \urlprefix\url{https://dx.doi.org/10.1103/physreve.74.061604}

\bibitem{Lauritsen1993}
Lauritsen K~B and Fogedby H~C 1993 {\em J. Stat. Phys.\/} {\bf 72} 189–205 \urlprefix\url{http://dx.doi.org/10.1007/BF01048046}

\bibitem{Takeuchi2017}
Takeuchi K~A 2017 {\em J. Phys. A: Math. Theor.\/} {\bf 50} 264006 \urlprefix\url{https://doi.org/10.1088/1751-8121/aa7106}

\bibitem{VaquerodelPino2025}
Vaquero~del Pino H and Cuerno R 2025 {\em Phys. Rev. Research\/} {\bf 7} \urlprefix\url{http://dx.doi.org/10.1103/y9b7-z6zq}

\bibitem{Kriecherbauer2010}
Kriecherbauer T and Krug J 2010 {\em J. Phys. A: Math. Theor.\/} {\bf 43} 403001 \urlprefix\url{https://dx.doi.org/10.1088/1751-8113/43/40/403001}

\bibitem{HalpinHealy2015}
Halpin-Healy T and Takeuchi K~A 2015 {\em J. Stat. Phys\/} {\bf 160} 794 \urlprefix\url{https://www.dx.org/10.1007/s10955-015-1282-1}

\bibitem{Carrasco2016}
Carrasco I~S~S and Oliveira T~J 2016 {\em Phys. Rev. E\/} {\bf 94} 050801(R) \urlprefix\url{https://www.dx.org/10.1103/PhysRevE.94.050801}

\bibitem{Carrasco2019}
Carrasco I~S~S and Oliveira T~J 2019 {\em Phys. Rev. E\/} {\bf 100} 042107 \urlprefix\url{https://www.dx.org/10.1103/PhysRevE.100.042107}

\bibitem{Young2015}
Young P 2015 {\em Everything You Wanted to Know About Data Analysis and Fitting but Were Afraid to Ask\/} (Heidelberg: Springer International Publishing)

\bibitem{Efron1982}
Efron B 1982 {\em The jackknife, the bootstrap, and other resampling plans\/} (Philadelphia, PA: SIAM)

\bibitem{Barreales2020}
Barreales B~G, Mel{\'{e}}ndez J~J, Cuerno R and Ruiz-Lorenzo J~J 2020 {\em J. Stat. Mech.: Theor. Exp.\/} {\bf 2020} 023203 \urlprefix\url{https://www.dx.org/10.1088/1742-5468/ab6a03}

\bibitem{Kallabis1997}
Kallabis H, Brendel L, Krug J and Wolf D~E 1997 {\em Int. J. Mod. Phys. B\/} {\bf 11} 3621–3634 \urlprefix\url{http://dx.doi.org/10.1142/S0217979297001829}

\bibitem{Michely2004}
Michely T and Krug J 2004 {\em Islands, {Mounds} and {Atoms}\/} (Berlin: Springer)

\bibitem{Banerjee2014}
Banerjee K, Shamanna J and Ray S 2014 {\em Phys. Rev. E\/} {\bf 90} ISSN 1550-2376 \urlprefix\url{http://dx.doi.org/10.1103/PhysRevE.90.022111}

\bibitem{Mal2016}
Mal B, Ray S and Shamanna J 2016 {\em Phys. Rev. E\/} {\bf 93} ISSN 2470-0053 \urlprefix\url{http://dx.doi.org/10.1103/PhysRevE.93.022121}

\bibitem{DasSarma2002}
Das~Sarma S, Chatraphorn P~P and Toroczkai Z 2002 {\em Phys. Rev. E\/} {\bf 65} \urlprefix\url{http://dx.doi.org/10.1103/PhysRevE.65.036144}

\bibitem{Nicoli2013}
Nicoli M, Cuerno R and Castro M 2013 {\em J. Stat. Mech. Theor. Exp.\/} {\bf 2013} P11001 \urlprefix\url{https://www.dx.org/10.1088/1742-5468/2013/11/p11001}

\bibitem{Alves2016}
Alves S~G and Ferreira S~C 2016 {\em Phys. Rev. E\/} {\bf 93} 052131 \urlprefix\url{https://link.aps.org/doi/10.1103/PhysRevE.93.052131}

\bibitem{Katzav2004}
Katzav E and Schwartz M 2004 {\em Phys. Rev. E\/} {\bf 70} \urlprefix\url{http://dx.doi.org/10.1103/PhysRevE.70.061608}

\bibitem{Haselwandter2006}
Haselwandter C~A and Vvedensky D~D 2006 {\em Phys. Rev. E\/} {\bf 73} \urlprefix\url{http://dx.doi.org/10.1103/PhysRevE.73.040101}

\bibitem{Meakin1998}
Meakin P 1998 {\em Fractals, scaling and growth far from equilibrium\/} vol~5 (Cambridge, UK: Cambridge University Press)

\bibitem{Oliveira2012}
Oliveira T~J, Ferreira S~C and Alves S~G 2012 {\em Phys. Rev. E\/} {\bf 85} 010601(R) \urlprefix\url{https://doi.org/10.1103/PhysRevE.85.010601}

\bibitem{Alves2013}
Alves S~G, Oliveira T~J and Ferreira S~C 2013 {\em J. Stat. Mech.: Theor. Exp.\/} {\bf 2013} P05007 \urlprefix\url{https://dx.doi.org/10.1088/1742-5468/2013/05/P05007}

\bibitem{Oliveira2013}
Oliveira T~J, Alves S~G and Ferreira S~C 2013 {\em Phys. Rev. E\/} {\bf 87} 040102 \urlprefix\url{https://link.aps.org/doi/10.1103/PhysRevE.87.040102}

\bibitem{Carrasco2018}
Carrasco I~S~S and Oliveira T~J 2018 {\em Phys. Rev. E\/} {\bf 98} 010102 \urlprefix\url{https://link.aps.org/doi/10.1103/PhysRevE.98.010102}

\bibitem{DeMauro2026}
de~Mauro G, Biroli M, Majumdar S~N and Schehr G 2026 {\em Phys. Rev. E\/} {\bf 113} ISSN 2470-0053 \urlprefix\url{http://dx.doi.org/10.1103/vmhz-6h5s}

\end{thebibliography}

\providecommand{\newblock}{}

\end{document}